\documentclass[a4paper,11pt]{article}
\pdfoutput=1 

\usepackage{jcappub} 

\usepackage[T1]{fontenc} 

\usepackage{subfigure} 

\usepackage{booktabs}
\usepackage{enumitem} 
\newcommand{\Cov}{{\rm Cov}}
\newcommand{\md}{{\rm d}}

\title{Robustness of dark energy phenomenology across different parameterizations}

\author{William J. Wolf,}
\affiliation{Astrophsyics, University of Oxford, United Kingdom}
\author{Carlos Garc\'ia-Garc\'ia,}
\author{Pedro G. Ferreira}

\emailAdd{william.wolf@stx.ox.ac.uk}
\emailAdd{pedro.ferreira@physics.ox.ac.uk}
\emailAdd{carlos.garcia-garcia@physics.ox.ac.uk}

\abstract{The recent evidence for dynamical dark energy from DESI, in combination with other cosmological data, has generated significant interest in understanding the nature of dark energy and its underlying microphysics. However, interpreting these results critically depends on how dark energy is parameterized. This paper examines the robustness of conclusions about the viability of particular kinds of dynamical dark energy models to the choice of parameterization, focusing on four popular two-parameter models: the Chevallier-Polarski-Linder (CPL), Jassal-Bagla-Padmanabhan (JBP), Barboza-Alcaniz (BA), and exponential (EXP) parameterizations. We find that conclusions regarding the viability of minimally and non-minimally coupled quintessence models are independent of the parameterization adopted. We demonstrate this both by mapping these dark energy models into the $(w_0, w_a)$ parameter space defined by these various parameterizations and by showing that all of these parameterizations can equivalently account for the phenomenology predicted by these dark energy models to a high degree of accuracy.}

\begin{document}
\maketitle
\flushbottom

\section{Introduction}

Understanding the nature of dark energy, alongside inflation and dark matter, is one of the three great mysteries of modern cosmology \cite{Frieman:2008sn, Peebles:2002gy, Joyce:2016vqv, Wolf:2024aeu, Ferreira:2025fpn}. Recent results from the DESI collaboration, in combination with data from supernovae and CMB experiments, have provided the first statistically significant evidence that dark energy may be dynamical \cite{DESI:2024mwx, Scolnic:2021amr, DES:2024tys, Rubin:2023ovl, Planck:2018vyg, ACT:2023kun}. Understandably, this has kicked off a wave of recent literature exploring these results and their implication for dark energy \cite{Wolf:2024eph, Wolf:2024stt, Ye:2024ywg, Tada:2024znt, Park:2024jns, DESI:2024kob, Dinda:2024kjf, Carloni:2024zpl, Wolf:2025jed, Wang:2024rjd, Mukherjee:2024ryz, Roy:2024kni, Wang:2024dka, Gialamas:2024lyw, Notari:2024rti, Wang:2024sgo, Wang:2024hwd, Giare:2024gpk, Dinda:2024ktd, Jiang:2024xnu, Ghosh:2024kyd, Luongo:2024fww, Reboucas:2024smm, Pang:2024qyh, Efstathiou_2024, Bhattacharya:2024hep, RoyChoudhury:2024wri, Arjona:2024dsr, Andriot:2024jsh, Wang:2024tjd, Berghaus:2024kra, Alestas:2024eic, Carloni:2024rrk, Chan-GyungPark:2024brx, Aboubrahim:2024cyk, Ye:2024zpk, Andriot:2024sif, Chudaykin:2024gol, Colgain:2024mtg, Gao:2024ily, Notari:2024zmi, Alfano:2024fzv, Luongo:2024zhc, Berbig:2024aee, Tiwari:2024gzo, Odintsov:2024woi, Colgain:2024ksa, Tang:2024lmo, Giare:2024ocw, Sakr:2025fay, Sakr:2025daj, Singh:2025seo, Ferrari:2025egk, Taule:2024bot, Colgain:2024xqj, Alfano:2025gie, Forconi:2025cwp, Jiang:2024viw, Yang:2025kgc, Huang:2025som, Dhawan:2024gqy, Hossain:2025grx, Lu:2025gki, Borghetto:2025jrk, Chakraborty:2025syu, Berti:2025phi, Yin:2024hba, Wang:2025ljj}. 

Much of this literature has focused on how significantly our interpretation of the data is influenced or determined by what are, to some extent, arbitrary choices in terms of how we parameterize the dark energy equation of state. For example, the commonly used Chevallier-Polarski-Linder (CPL) parameterization, $w(a) = w_0 + w_a \left(1-a\right)$, assumes that the dark energy equation of state evolves linearly with the scale factor and is given in terms of two parameters: the value of the equation of state now, $w_0$, and its time derivative, $w_a$ \cite{Chevallier:2000qy, Linder:2002et}. Yet, recent work has pointed to certain issues or ambiguities that can arise when using it, such as its sensitivity to the redshifts probed by the data and the potential for misleading inferences when it is extrapolated outside the ranges of the data constraining it \cite{Wolf:2023uno, Cortes:2024lgw, Shlivko:2024llw, Wolf:2024eph}. Of course, there are no \textit{a priori} reasons that guarantee that a linear form for the equation of state is sufficient to capture the relevant physics of dark energy and physicists have proposed other similar ways of parameterizing the dark energy equation of state in terms of two parameters with more intricate functional forms \cite{Pan:2019brc, Jassal:2005qc, Barboza:2008rh}.\footnote{There are a tremendous variety of other options as well besides the most popular two parameter models considered here, including other two parameter models as well as one and three parameter models. See e.g.~\cite{Alho:2024qds, Payeur:2024dnq, Akthar:2024tua, Mukherjee:2016eqj, Feng:2012gf, Taylor:2024whh, Singh:2023ryd, Efstathiou:1999tm, Cooray:1999da, Gong:2005de, Li:2020ybr, Fikri:2024klc, Afroz:2024lou, Chan-GyungPark:2025cri, Shlivko:2025fgv} and references therein for a small sample of this large literature.} While recent studies have found that these too show a preference for dynamical dark energy \cite{Giare:2024gpk, Zheng:2024qzi, Pourojaghi:2024tmw, DESI:2025fii}, these proposals do not obviously resolve the ambiguities that have been pointed out in the case of the CPL parameterization. And finally, physicists have pursued more model agnostic methods, which have led to conflicting results depending on the approach adopted, with some favoring dynamical dark energy and others favoring $\Lambda$CDM (see e.g.~\cite{Dinda:2024kjf, Dinda:2024ktd, Mukherjee:2024ryz, DESI:2024kob, Ye:2024ywg}). 

Another long-standing theme of the dark energy literature is assessing what data constraints on parameterized models of dark energy imply about the microphysics of dark energy. In other words, can we use constraints on these parameters to infer what kinds of microphysical or field-theoretic models of dark energy are favored or viable \cite{Linder:2006sv, Linder:2007wa, Caldwell:2005tm, Scherrer:2005je, Wolf:2023uno, Shlivko:2024llw, Linder:2024rdj, Lewis:2024cqj, Wolf:2024stt, Wolf:2024eph}? There are a number of ways of going about this, but in the literature it has been common to generate such microphysical models and then fit one of these parameterizations either to their equations of state or to their Hubble rates, over some recent range of redshifts where dark energy is believed to be most relevant, and to then take this as a suitable proxy for how well a dark energy model aligns with the data (suitably interpreted in terms of one of the parameterizations). More recently, \cite{Wolf:2024eph} developed a procedure to translate dark energy models into CPL parameters by directly using the cosmological data. Essentially, the idea is to find the best fitting $(w_0, w_a)$ values that reproduce the specific cosmological observables probed in experiments and predicted by a dark energy theory, while using the known errors and covariances of the data in determining the best fit (see also \cite{Garcia-Garcia:2019cvr, Traykova:2021hbr} for an earlier iteration). In \cite{Wolf:2024eph}, this was used to show that thawing quintessence does not share significant overlap with the regions of parameter space favored by the CPL model, while in \cite{Wolf:2024stt} it was shown that a non-minimally coupled dark energy model with a recent phantom crossing does share significant overlap with the data. Directly comparing the CPL parameterization and these dark energy models with respect to their $\chi^2$ statistics for fitting the full complement of data then validated the intuitions developed in mapping these microphysical models into the CPL parameter space.  

This paper brings these strands of the literature together, focusing on the most popular two-parameter dark energy parameterizations which include the CPL, Jassal-Bagla-Padmanabha
(JBP), Barboza-Alcaniz (BA), and exponential (EXP) parameterizations. The questions we seek to answer are:
\begin{enumerate}[label=(\roman*)]
    \item Are the conclusions about which kinds of microphysical models are being favored by the data robust to these different assumptions that parameterize dark energy evolution? For example, would adopting a different parameterization change the conclusion that canonical, minimally coupled quintessence fields are a poor description of the data (while certain kinds of non-minimally coupled models fair much better) when the data is interpreted through the lenses of different parametric forms of the equation of state?
    \item Are these parameterizations in fact adequate descriptions of dark energy physics? After all, information is certainly lost when we assume that microphysical dark energy models can be parameterized in such simple ways and there is no guarantee that these parameterizations are descriptive enough of microphysical dark energy phenomenology for us to trust the inferences we make when adopting them.
\end{enumerate}
As we shall see, we will answer in the affirmative to both of these questions. That is, at least with respect to all of these commonly used two parameter models, our conclusions about the viability of certain kinds of microphysical dark energy models are robust. And furthermore, this is precisely because all of the parameterizations perform similarly well in terms of describing microphysical dark energy phenomenology. That is, they all are able to offer functionally equivalent descriptions of the cosmological observables that the dark energy models would predict and these descriptions are a sufficiently accurately approximate to facilitate a comparison of how well these dark energy models will perform with respect to the actual cosmology data; consequently, our conclusions are robust regardless of which parameterization is adopted. We will explore this in the case of general models of both minimally and non-minimally coupled quintessence. So while there are many various motivations to consider distinct ways of parameterizing dark energy, it is the case that these two parameter models are at present sufficient and there is no significant need to move to more complicated descriptions. However, it is of course important to be aware of their shortcomings and to not over-interpret them. 

The paper proceeds as follows. Section \eqref{Sec:DEparam} introduces the most commonly used parameterizations for the dark energy equation of state. Section \eqref{Sec:DEmodels} introduces our models of minimally and non-minimally coupled quintessence. Section \eqref{Sec:data} discusses the data used in the analsis and the procedure used for mapping dark energy models into $(w_0, w_a)$ by using the data directly. Section \eqref{Sec:phenom} demonstrates that all of the parameterizations considered here do a good job of reproducing the predicted cosmological observables for the microphysical dark energy models considered here. Section \eqref{Sec:plane} maps the dark energy models considered here into the $(w_0, w_a)$ parameter space for all of the parameterizations considered here. Section \eqref{Sec:conclude} concludes.

\section{Parameterizing Dark Energy}\label{Sec:DEparam}

We do not measure the equation of state directly. Rather we measure cosmological quantities that are sensitive to $H(z)$ at different redshift epochs. Thus, to properly compare with cosmological data, we must fit actual measurements that are sensitive to $H(z)_{obs}$ to theoretical models of $H(z)$. 
\begin{equation}
H^2(a) = H_0^2 \left[ \Omega_{\mathrm{r}} a^{-4} +  \Omega_{\mathrm{m}} a^{-3} + \Omega_{\mathrm{de}} e^{3 \int_a^1\left(1+w_{\text {de }}\right) \mathrm{d} \ln \mathrm{a}} \right]
\end{equation}
In order to do this efficiently, it is common practice to parameterize $w(a)$ in terms of 2 numbers, ($w_0$, $w_a$). These parameters ($w_0$, $w_a$) can then be used to model the behavior of $H(z)$ given an equation of state that evolves in that way. Below, we briefly review the most common ways of doing so.

\subsection{Chevallier-Polarski-Linder parameterization}
The standard way of parametrising time-varying dark energy is given by the Chevallier-Polarski-Linder (CPL) form \cite{Chevallier:2000qy, Linder:2002et},
\begin{equation}\label{eq:cpl}
    w(a) = w_0 + w_a \left(1-a\right),
\end{equation}
where $w_0$ is the value of the equation of state now and $w_a$ captures the time variation of the equation of state.
The parameterization is essentially conceived of as a Taylor expansion of the equation of state at recent times ($a\simeq1$), with the linear term capturing the leading order temporal variation of the equation of state. 
\begin{equation}
    \frac{dw}{da} = -w_a.
\end{equation}

In reality though and as discussed above, we do not measure the equation of state directly and must instead determine how such an equation of state will influence $H(z)$ to facilitate a one-to-one comparison with the data. One finds that under this parameterization of dark energy, the contribution to the Friedmann equation is:
\begin{equation}
H^2(a)=H_0^2\left[\Omega_{\mathrm{m}} a^{-3}+\left(1-\Omega_{\mathrm{m}}\right) e^{3 w_a(a-1)} a^{-3\left(1+w_0+w_a\right)}\right].
\end{equation}
Actually determining ($w_0$, $w_a$) then involves finding the best fit ($w_0$, $w_a$) given the particular observations considered in a survey (to be discussed further in Section \eqref{Sec:data} and Appendix \eqref{sec:compressed}).

While this parameterization is by far the most common, recent research has raised some questions about its utility and urged caution when interpreting it (e.g.~\cite{Cortes:2024lgw, Wolf:2023uno, Shlivko:2024llw, Wolf:2024eph}). Namely, if dark energy evolves non-linearly, there will not be a unique representation of such models in the ($w_0$, $w_a$) plane as the best linear fit will be sensitive to which redshifts are probed. To this point, it has been shown that using this parameterization can create the false impression that certain dark energy models are ``phantom'', when in reality they are non-phantom throughout their entire evolution but only appear to be phantom because the best fit ($w_0$, $w_a$) parameters are being extrapolated outside the redshift epochs which were considered in the fit. Thus, it is worth considering other ways of parameterizing the dark energy equation of state.

\subsection{Exponential parameterization}
One can also consider an exponential form of the parameterization \cite{Dimakis:2016mip, Pan:2019brc, Najafi:2024qzm},
\begin{equation}\label{eq:exp}
w(a) = w_0 + w_a (e^{(1-a)}-1).
\end{equation}
Upon taking a Taylor expansion of this exponential parameterization, we can see that it can be viewed as extended version of the CPL parameterization as it allows for corrections beyond linear order. The advantage here is that it may very well be the case that a purely linear parameterization is too blunt of an instrument to accurately capture the dynamical effects of dark energy over the entire expansion history of the universe. Thus, considering a parameterization of this form is a natural extension that allows us to explore whether next to leading order terms might be important. Following \cite{Najafi:2024qzm}, we will go to second order and adopt the following form of the exponential parameterization,
\begin{equation}\label{eq:exp}
w(a)=w_0+w_a\left[(1-a)+\frac{1}{2!}\left(1-a\right)^2\right],
\end{equation}
which allows for deviations from the base CPL scenario and gives the following contribution to the Friedmann equation (written in terms of the familiar CPL form plus the additional correction coming from the next order term):
\begin{equation}
H^2(a)=H_0^2\left[\Omega_{\mathrm{m}} a^{-3}+\left(1-\Omega_{\mathrm{m}}\right) e^{3 w_a(a-1)} a^{-3\left(1+w_0+w_a\right)}e^{3 w_a\left(\frac{1}{4}a^2+\frac{1}{2}a-\frac{3}{4}\right)}a^{-\frac{3}{2}w_a} \right].
\end{equation}

\subsection{Jassal-Bagla-Padmanabha parameterization}
If one wants to depart further from the CPL parameterization, we can consider the Jassal-Bagla-Padmanabha (JBP) parameterization \cite{Jassal:2005qc},
\begin{equation}\label{eq:jbp}
w(a)=w_0+w_a \left(a-a^2\right).
\end{equation}
This parameterization is interesting in the sense that it adopts the perspective that if dark energy is an insignificant dynamical factor before $z \gtrsim 1$, one should instead focus on the regime $z \lesssim 1$. Thus, the JBP parameterization evolves non-linearly and has $w(0)=w(1)=w_0$. The contribution to the Friedmann equation is given by:
\begin{equation}
H^2(a)=H_0^2\left[\Omega_{\mathrm{m}} a^{-3}+\left(1-\Omega_{\mathrm{m}}\right) e^{\frac{3}{2} w_a (1-a)^2} a^{-3\left(1+w_0\right)}\right].
\end{equation}

\subsection{Barboza-Alcaniz parameterization}
We will also consider the Barboza-Alcaniz (BA) parameterization \cite{Barboza:2008rh},
\begin{equation}\label{eq:BA}
w(a) =w_0+w_a \left(\frac{1-a}{a^2+(1-a)^2}\right).
\end{equation}
Like the JBP parameterization, it departs from strict linearity and allows for more complicated evolution in the equation of state. However, it also explicitly aims to cover the entire expansion history of the universe as it is a well-defined function over all redshifts without enforcing any particularly suspect behavior at high redshifts, and has been argued to do a better job of minizing errors at low redshift than the CPL or JBP models \cite{Colgain:2021pmf}. 
The contribution to the Friedmann equation is:
\begin{equation}
H^2(a)=H_0^2\left[\Omega_{\mathrm{m}} a^{-3}+\left(1-\Omega_{\mathrm{m}}\right) \left(\frac{a^2+(1-a)^2}{a^2}\right)^{\frac{3}{2}w_a}a^{-3\left(1+w_0\right)}\right].
\end{equation}.

\section{Dark Energy Models}\label{Sec:DEmodels}

\subsection{Thawing quintessence with minimal coupling}

Quintessence refers to dark energy driven by a single scalar field, $\varphi$, which is minimally coupled to gravity and has a canonical kinetic energy. The action for the theory is:
\begin{equation}\label{eq:TQ}
S=\int d^4 x \sqrt{-g}\left[\frac{1}{2} M_{\mathrm{\rm P}}^2 R-\frac{1}{2} g^{\mu \nu} \partial_\mu \varphi \partial_\nu \varphi-V(\varphi)\right]+S_{\rm m}.
\end{equation}
$M_{\rm P}$ is the reduced Planck mass, $g$ is the determinant of the metric $g_{\mu\nu}$, $R$ is the Ricci curvature scalar, $V(\varphi)$ is the scalar field potential, and $S_{\rm m}$ is the action for matter. 

For quintessence, the dark energy equation of state $w(a)$ is given by:
\begin{equation}\label{DEeq}
 w(a) \equiv \frac{P_{\varphi}}{\rho_{\varphi}}=\frac{\frac{\dot{\varphi}^2}{2}-V(\varphi)}{\frac{\dot{\varphi}^2}{2}+V(\varphi)},
\end{equation}
where $\rho_\varphi$ and $P_\varphi$ are the scalar field energy density and pressure, respectively.
The potential $V(\varphi)$ is the most significant factor in determining the evolution of $w(a)$ and the phenomenological behavior of dark energy. When $V(\varphi)$ dominates over the kinetic energy of the scalar field $\dot{\varphi}^2/2$, the equation of state approaches that of a cosmological constant $w(a) \simeq -1$. However, $w(a)$ will change over time due to the evolution of the scalar field $\varphi$. The dynamics of this evolution are governed by the scalar field's equation of motion in a Friedmann background: 
\begin{equation}\label{ScalarEOM}
    \ddot{\varphi} + 3 H \dot{\varphi} + V'(\varphi) = 0,
\end{equation}
where $V'(\varphi) \equiv dV/d\varphi$, and $H$ represents the expansion rate of the Universe, determined by the first Friedmann equation:
\begin{equation}\label{Friedmann}
H^2\equiv 
\left(\frac{\dot{a}}{a}\right)^2=\frac{1}{3M^2_{\rm P}} \left[\frac{1}{2} \dot{\varphi}^2 + V(\varphi)+\rho_{\rm m}\right],
\end{equation}
with $\rho_{\rm m}$ the energy density of the ordinary matter, respectively. The Hubble parameter today is $H_0 = 100 h$ km s$^{-1}$ Mpc$^{-1}$, with $h \simeq 0.67$.

As the data favors so called ``thawing'' models of dark energy, for which the equation of state grows larger as it evolves ($dw/da>0$), we will focus on a general model of quintessence which can be understood to capture the entire phenomenology of $w(a)$ associated with thawing quintessence. In \cite{Wolf:2023uno, Wolf:2024eph, Wolf:2025dss}, it has been argued that this can be accomplished with a particular simple model given by an energy scale $V_0$ and a quadratic term with a mass $m^2$,
\begin{equation}\label{potential}
    V (\varphi) = V_0 \pm \frac{1}{2}m^2\varphi^2,
\end{equation}
where throughout this paper, we will give $m^2$ in units of $(H_0/h)^{2}$ and $V_0$ in units of $M^2_{\rm P}(H_0/h)^{2}$. This model has been implemented in \texttt{hi\_class} \cite{hi_class_1,hi_class_2}, where the initial conditions are chosen to generically set $\dot{\varphi}_{ini}= 0$ and $\varphi_{ini}$ is tuned to satisfy the equations of motion while the other parameters are varied (see \cite{Wolf:2024eph} for an analytic expression for $\varphi_{ini}$ in terms of model and cosmological parameters).

This model can capture the entire phenomenology of thawing quintessence because, when we have a positive mass term in Eq.~\eqref{potential}, it behaves as ``slow-roll'' thawing quintessence \cite{Scherrer:2007pu}, which has been extensively studied and showed to converge to a universal phenomenological behavior given by an approximately linear evolution in the equation of state. This behavior is well-described by $dw/da|_{a=1} \approx 1.5 [1+ w(a=1)]$ (e.g.~\cite{Scherrer:2007pu, Marsh:2014xoa, Garcia-Garcia:2019cvr, Wolf:2023uno, Scherrer:2015tra, Linder:2015zxa}). When the mass is negative, this corresponds to ``hilltop'' quintessence \cite{Dutta:2008qn, Chiba:2009sj, Shlivko:2024llw, Wolf:2023uno, Borghetto:2025jrk}, which has much more varied dynamical behavior. If $\left|-m^2\right|$ is small, the resulting $w(a)$ approximates the linear behavior of the positive branch. On the other hand, if $\left|-m^2\right|$ is large, $w(a) \simeq -1$ for much of cosmic history, but can evolve very rapidly and non-linearly at more recent times depending on the choice of model parameters, allowing it to ``sweep'' over the $(w_0, w_a)$ parameter space. In addition to the above arguments, we can also understand why this model is able to subsume the phenomenology of thawing quintessence in the following way. As any arbitrary analytic potential can be expand in a Taylor series (e.g.~\cite{Dutta:2008qn, Wolf:2023uno, Chiba:2009sj}), the potential given in Eq.~\eqref{potential} can be seen as the leading order terms for a huge variety of scalar field potentials that broadly fall under either the ``slow-roll'' and ``hilltop'' types; and in the regime of field space for which quintessence can play the role of cosmological dark energy, these are the terms that will be relevant.

We refer to this model as \textit{Thawing Quintessence} given we can interpret so many of these distinct models simply in terms of an energy scale and mass, and \cite{Wolf:2024eph} determined constraints on the microphysics of the model given by Eq.~\eqref{potential}. The findings heavily favored negative masses as these parameter choices lead to a more sharp evolution in the equation of state. Given the dynamical features of the negative mass branch discussed above, this is perfectly sensible. As discussed in  \cite{Wolf:2023uno, Wolf:2024eph, Shlivko:2024llw, Tada:2024znt}, the data pushes us into the steeper regions of the CPL ($w_0$, $w_a$) plane and this requires sharper temporal variation in $w_a$ than is given by the models which evolve according to  $dw/da|_{a=1} \approx 1.5 [1+ w(a=1)]$. However, as discussed in \cite{Wolf:2024eph}, while these models do improve the fit to the data over $\Lambda$CDM, once one uses various information theoretic accounts for the introduction of new parameters, the improvement in fit they offer over $\Lambda$CDM is not as compelling and the evidence for them is scant. More and better data is required before any firm verdicts can be reached.

\subsection{Thawing quintessence with non-minimal coupling}
Another possibility for dark energy is to introduce a non-minimal coupling into the action:
\begin{align}\label{eq:NMTQ}
S=\int d^4 x\sqrt{-g}\left[\frac{1}{2}\left(M^2_{\rm P}-\xi\varphi^2\right)R-\frac{1}{2}\partial_\mu\varphi\partial^\mu\varphi - V(\varphi) \right] + S_m,
\end{align}
where $\xi$ is the dimensionless coupling coefficient between the scalar field $\varphi$ and gravity $R$. This introduces a modification into the gravitational sector (while keeping the matter sector of minimally coupled quintessence the same)\footnote{Other kinds of non-minimal couplings such as a direct coupling between the scalar field and dark matter are also possible (e.g.~\cite{Gomez-Valent:2020mqn, Goh:2022gxo, Amendola:1999er, Giare:2024smz, Chakraborty:2025syu,Khoury:2025txd}).
}, which changes the corresponding scalar field equation of motion as well as the standard Friedmann equation from GR:
\begin{equation}
\ddot{\varphi}+3 H \dot{\varphi}+V'(\varphi)+6\left(2+\frac{\dot{H}}{H^2}\right) \xi H^2 \varphi=0,
\end{equation}

\begin{equation}
    H^2 = \frac{1}{3 M_{\text{\rm P}}^2 \left(1 - \xi \varphi^2 \right)} \left[ \frac{1}{2} \dot{\varphi}^2 + V(\varphi) + 3 H \xi \varphi \dot{\varphi} + \rho_m \right].
\end{equation}

This leads to a different expression for the equation of state given by the following:
\begin{equation}
w(a) \equiv \frac{-2 \dot{H}-3 H^2-P_{\mathrm{m}}}{3 H^2-\rho_{\mathrm{m}}}.
\end{equation}
Interestingly, this modification can lead to a stable phantom crossing at recent times, which seems to be preferred by the expansion history data. As discussed in \cite{Wolf:2024stt, Ye:2024ywg, Ye:2024zpk, Chudaykin:2024gol, Tiwari:2024gzo, Ferrari:2025egk, Taule:2024bot}, dark energy models with a non-minimal coupling 
can improve the fit to the data over $\Lambda$CDM or quintessence. In particular, models that combine a non-minimal coupling like that in Eq.~\eqref{eq:NMTQ} with a ``hilltop'' potential, 
\begin{equation}
    V(\varphi) = V_0\left(1-\beta e^{-\lambda \varphi }\right),
\end{equation}
as was done in \cite{Wolf:2024stt} can have a recent phantom crossing along with rapid dynamical evolution and display a significant improvement to the fit of the data over $\Lambda$CDM or quintessence.

In this paper, we instead adopt a more ``agnostic'' potential of the following form:
\begin{equation}\label{nm:potential}
    V (\varphi) = V_0 \left(1 + \beta \varphi + \frac{1}{2}m^2\varphi^2\right),
\end{equation}
where, as before, each of these terms can be thoughts of as the leading order terms in a Taylor expansion and the coefficients can be positive or negative. This model has been implemented in \texttt{hi\_class} \cite{hi_class_1,hi_class_2}, where the initial conditions are chosen such that $\varphi_{ini}=0$ and $\dot{\varphi}_{ini}\simeq 0$. We tune $V_0$ to satisfy the equations of motion, finding that $V_0 \simeq 3 H_0^2 \Omega_{\varphi}$ is a robust starting point for the numerical solvers. Similar to before, the units of $V_0$ are given in $M^2_{\rm P} (H_0/h)^{2}$, the units of $\beta$ are given in $M_{\rm P}(H_0/h)^{2}$, and the units of $m^2$ are given in $(H_0/h)^{2}$. Of course, as a modified gravity model, this model can suffer from instabilities (see e.g.~\cite{Wolf:2019hzy, Bellini:2014fua}). In this paper we will be primarily concerned with predictions for background quantities, but we use \texttt{hi\_class} to perform stability checks and ensure viability of all the parameter choices considered here by computing linear cosmological perturbations for each model.  

In analogy with thawing quintessence above, we can refer to this as \textit{Non-Minimal Thawing Quintessence} as it will phenomenologically capture the behavior of dark energy models with this kind of non-minimal coupling and an arbitrary potential. For example, it captures the behavior of non-minimal models with a simple exponential potential and with a hilltop potential which have been studied in \cite{Wolf:2024stt, Ye:2024ywg, Ye:2024zpk}, as again this form can be seen as corresponding to a general Taylor expansion in the potential. In \cite{Wolf:2024stt}, it was demonstrated that such models can saturate much of the thawing region of the CPL $(w_0, w_a)$ plane. See also \cite{Wolf:2025jed} for a full Bayesian analysis where it is found that this type of non-minimally coupled dark energy model is strongly favoured over $\Lambda$CDM. However, in this paper, we will focus on the phenomenology and whether or not the aforementioned dark energy parametrizations are equipped to capture it. 

\subsection{Projecting dark energy models onto the $(w_0, w_a)$ plane}\label{Sec:projection}
We follow \cite{Garcia-Garcia:2019cvr, Wolf:2024eph} to compress the phenomenology of the scalar field models onto the $(w_0, w_a$) plane. In short, given a set of cosmological and dark energy parameters, we generate the set of observables predicted by the dark energy models that correspond to the data measurements that have been taken at particular redshifts in the various surveys considered, apply the data covariance (i.e.\ errors), and find the best fitting $(w_0, w_a)$ parameters; i.e., we minimize
\begin{equation}\label{chi2}
\chi^2 = \left(\mathcal{O}^{({w_0, w_a})} - \mathcal{O}^{\varphi} \right)^T \left(\frac{\mathcal{O}^{\rm data}}{\mathcal{O}^{\varphi}}\right)^T {\Cov}^{-1} \frac{\mathcal{O}^{\rm data}}{\mathcal{O}^{\varphi}} \left(\mathcal{O}^{({w_0, w_a})} - \mathcal{O}^{\varphi} \right),
\end{equation}
where, $\mathcal{O}^{({w_0, w_a})}$ and $\mathcal{O}^{\varphi}$ are the observables computed with the CPL parameterization or the dark energy model respectively, the factors $\mathcal{O}^{\rm data} / \mathcal{O}^{\varphi}$ ensure that we fit the observables using the data measurements' relative errors, and ``Cov'' represents the associated covariance. See Appendix \eqref{sec:compressed} for a detailed discussion of the methods used and validation of this procedure for compressing the datasets.

In order to explore the parameter space, we generate viable models by sampling the cosmological and dark energy parameters in a Latin hypercube. To avoid wasting computational time in observationally excluded regions, we sample around current data constraints for the background cosmological parameters: $\Omega_{\rm m}\in [0.27, 0.34]$ and $h \in [0.64, 0.72]$. These values enclose the 3$\sigma$ ranges around combined CMB, DESI BAO, and SNe (for both Pantheon+ and DES-Y5) data used in \cite{DESI:2024mwx} and here\footnote{Note that we do not need to enlarge the range of $h$ to encompass the \cite{Riess:2021jrx} value since we are not using calibrated SNe and, therefore, the constraints on $h$ are mainly driven by Planck.}.

For thawing quintessence, we break the sampling up into the positive $m^2$ branch and the hilltop/negative $m^2$ branch. Essentially, the reasons for this are that the hilltop branch has $m^2$ unbounded: as one moves to ever larger masses, one can fine-tune $\varphi_{ini}$ ever closer to the top of the hill, which allows a range of masses much more extended than for the positive mass branch (see Fig.~(5) in \cite{Wolf:2024eph}).  As a consequence, it is more efficient to sample each branch separately. For the hilltop branch, we generate $\simeq$ 15,000 viable models in the ranges $m^2 \in [-50.0, 0]$ and $V_0 \in [0.9, 1.4]$, while for the positive mass branch we generate $\simeq$ 5,000 models in the ranges $m^2 \in [0.0, 2.5]$ and $V_0 \in [-0.5, 0.5]$.  

For non-minimal thawing quintessence, we generate $\simeq$ 10,000 models in the ranges $\xi \in [1.5, 3.0]$, $\beta \in [0.0, 10.0]$, and $m^2 \in [-10.0, 2.0]$. The flexibility of this model; however, allows it to cover a very wide range of behaviors for $w(a)$, including large parts of the phantom region today ($w_0 < -1$) and huge swaths of the thawing region ($w_a < 0$). However, for the purposes of this paper, the range of parameters was chosen to reproduce the slice of parameter space that most closely corresponds to the current constraints on the equation of state. In all cases considered here, we have validated that the number of models sampled is sufficient as the results do not change with the addition of more models.

\section{Data and likelihood}\label{Sec:data}

In this article, we use the following datasets to constrain the dark energy parameterizations in Section \eqref{Sec:DEparam}:
\begin{itemize}
    \item \textbf{DESI DR1 BAO:} We use the DESI DR1 BAO measurements, which consist of a set of twelve measurements of the BAO scale, from 6 million galaxies in the range of redshifts $0.1 < z < 4.2$. The cosmological quantities measured are the angular diameter distance, $D_M$, the Hubble parameter, $D_H = c/H$, and the angle-averaged quantity $D_V = (z D_M^2 D_H)^3$ \cite{DESI:2024mwx}.
    \item \textbf{Pantheon+ SNe:} We use the Pantheon+ sample, which is comprised of 1550 SNe Ia in the range $0.001<z<2.26$ \cite{Scolnic:2021amr}.
    \item \textbf{DES-Y5 SNe:} We use the DES-Y5 sample, which is comprised of 1635 light curves from 1550 SNe Ia in the range $0.10<z<1.13$  and 194 low-redshit SNe Ia in the range $0.025<z<0.10$ \cite{DES:2024tys}.
    \item \textbf{Planck 2018:} We use the auto- and cross-correlation of the CMB temperature and $E$-mode polarization fields. We use the low-$\ell$ power spectra, obtained with the ``Commander'' component separation algorithm, in the range $2 \leq \ell \leq 29$ and the high-$\ell$ \texttt{plik} likelihood covering $30 \leq \ell \leq 2508$ for the temperature auto-correlation ($TT$) and $30 \leq \ell \leq 1996$ for the $TE$ and $EE$ components \cite{Aghanim:2019ame}.
    \item \textbf{ACT DR6 lensing:} We use CMB lensing from ACT DR6, consisting of the ACT CMB reconstructed lensing power spectra between the scales $40 < L < 763$ \cite{ACT:2023dou, ACT:2023kun}. 
\end{itemize}
For the model projection onto the $({w_0, w_a})$ plane we use compressed versions of the CMB and SNe data: we compress these likelihoods onto a few background quantities as explained in Appendix \eqref{sec:compressed}. This simplifies the minimization without any impact on the results.

\begin{table}[t]
\centering
\begin{tabular}{|c|c|}
\hline
\multicolumn{2}{|c|}{\textbf{Cosmological Parameters}} \\
\hline
$\omega_b$ & $\mathcal{U}[0.005, 0.1]$ \\
$\Omega_m$ & $\mathcal{U}[0.01, 0.99]$ \\
$H_0 \ [\text{km/s/Mpc}]$ & $\mathcal{U}[20, 100]$ \\
$n_s$ & $\mathcal{U}[0.8, 1.2]$ \\
$\ln 10^{10} A_s$ & $\mathcal{U}[1.61, 3.91]$ \\
$\tau$ & $\mathcal{U}[0.01, 0.8]$ \\
\hline
\multicolumn{2}{|c|}{\textbf{Dark Energy Parameters}} \\
\hline
$w_0$ & $\mathcal{U}[-3.0, 1.0]$ \\
$w_a$ & $\mathcal{U}[-3.0, 2.0]$ \\
\hline
\end{tabular}
\caption{Prior distributions used in the cosmological parameter inference. $\mathcal{U}(a, b)$ stands for an uniform distribution in the range $[a, b]$.}\label{priors}
\end{table}

We constrain the cosmological parameters and $({w_0, w_a})$ using the Metropolis-Hasting algorithm \cite{metropolismc, hastingsmc, Lewis:2002ah, Lewis:2013hha} and data likelihoods as implemented in \texttt{Cobaya}. We sample the cosmological parameters $\{\omega_b, \Omega_m, H_0, n_s, \ln 10^{10} A_s, \tau\}$, together with $(w_0, w_a)$, with the priors shown in  Table \eqref{priors} (while keeping the neutrino mass fixed to \(\sum m_\nu = 0.06\,\mathrm{eV}\) throughout). We use the Boltzmann code \texttt{hi\_class} to compute the observables and assess convergence with the Gelman-Rubin criterion \cite{gelmanrubin}, requiring $R-1 < 0.02$.

\section{Results}
In this section we show the constraints on the $(w_0, w_a)$ parameter space for the different parametrizations and compare them with the projections of the scalar field models into their parameters to assess their fit to the data. We show all results for the combination of DESI DR1 BAO, CMB and SNe Ia data (Pantheon+ or DES-Y5 data).

\subsection{Dark energy models in the $(w_0, w_a)$ plane}\label{Sec:plane1}

\begin{figure}[t]
  \centering
       \includegraphics[width=0.48\textwidth]{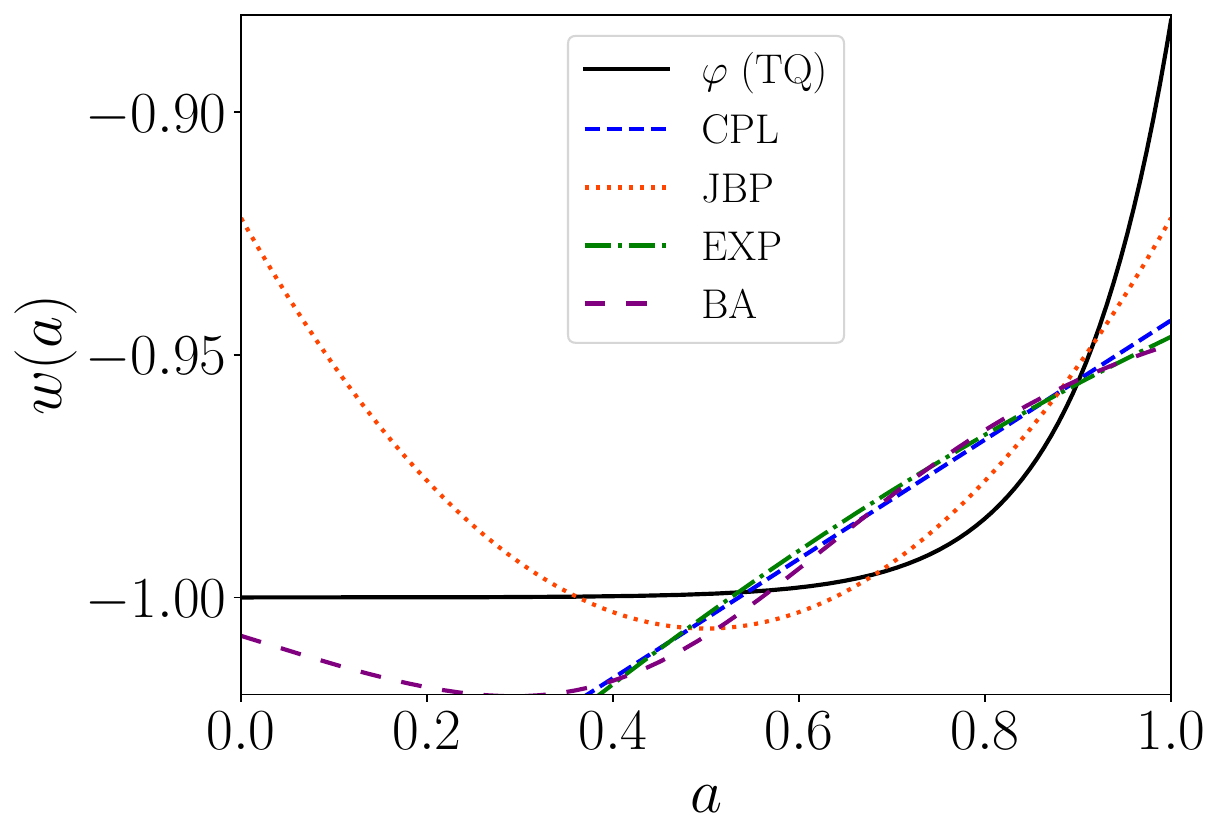}
       ~
       \includegraphics[width=0.48\textwidth]{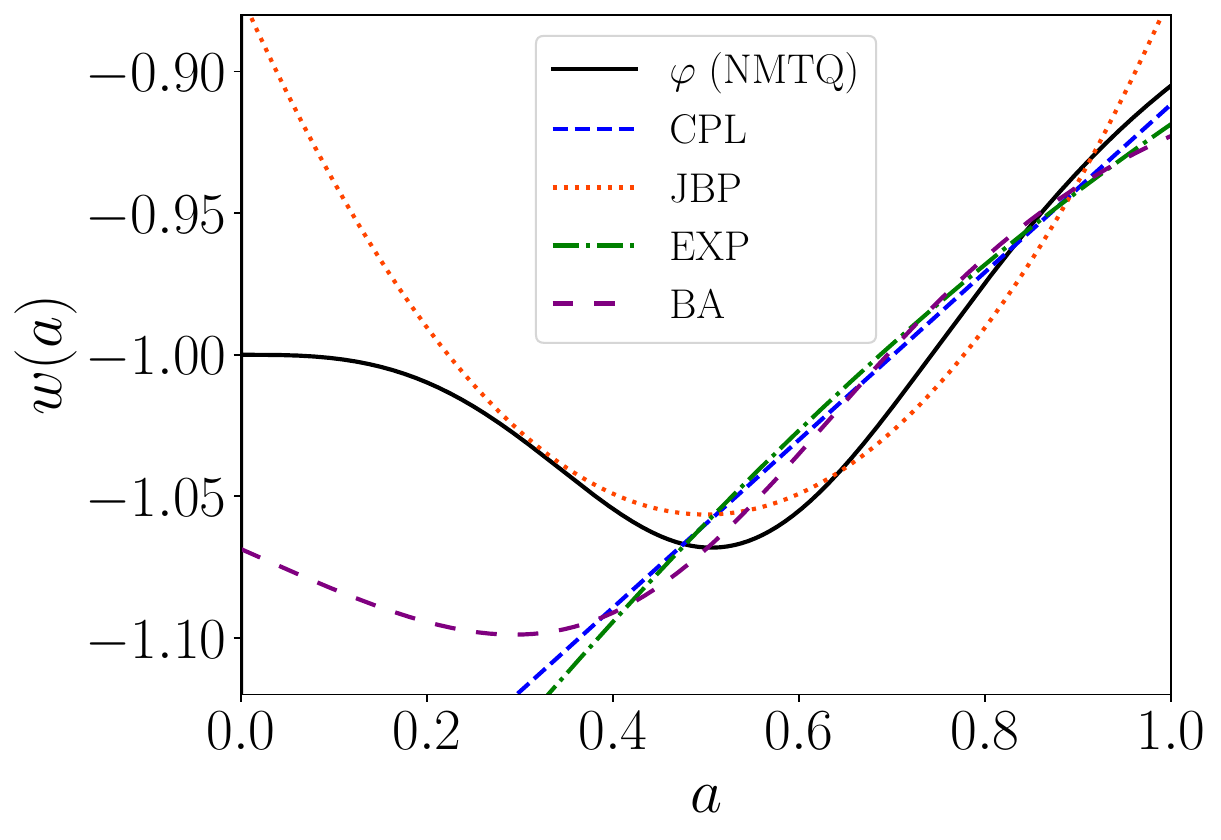}

  \caption{Equation of state, $w(a)$, for representative dark energy models from both thawing quintessence (left) and non-minimal thawing quintessence (right), alongside the corresponding best fit CPL, JBP, EXP, and BA models when these are determined by fitting directly to the compressed BAO, SNe, and CMB data using Eq.~\eqref{chi2}.}
  \label{fig:fits}
\end{figure}
We start by showing how the different parametrizations represent the equation of state. Fig.~\eqref{fig:fits} shows the equation of state for a representative thawing quintessence and a representative non-minimal thawing quintessence model and the corresponding best fitting parametrizations. The dark energy models have cosmological parameters $\Omega_m = 0.315$, $H_0 = 67.5$ and dark energy parameters $(V_0, m^2)=(0.975, -16.0)$ for thawing quintessence and $(\xi, V_0, \beta, m^2)=(1.5, 0.83, 2.0, -2.0)$, for the non-minimal case. The best-fitting parameters for the thawing quintessence model are: CPL: $(w_0, w_a) = (-0.94, -0.12)$, JBP: $(w_0, w_a) = (-0.92, -0.34)$, EXP: $(w_0, w_a) = (-0.95, -0.09)$, and BA: $(w_0, w_a) = (-0.95, -0.06)$; and for the non-minimal model: CLP: $(w_0, w_a) = (-0.91,  -0.30)$, JBP: $(w_0, w_a) = (-0.87, -0.73)$, EXP: $(w_0, w_a)= (-0.92, -0.23)$, and BA: $(w_0, w_a) = (-0.92, -0.15)$. While the various dark energy parameterizations all produce reasonable looking fits (again determined by fitting the observables directly) when plotted against the equation of state, they can also be misleading when extrapolated over all of cosmic history as the results will depend on the redshifts probed by the data and their associated uncertainties (see \cite{Wolf:2023uno, Wolf:2024stt, Shlivko:2024llw, Wolf:2024eph} for more related discussion). That is, $w_0$ often differs substantially from $w(a=1)$ (even though it is often interpretted as being the value of the equation of state today) because the best fit $(w_0, w_a)$ is determined by fitting over a range of observations mostly from the distant past, meaning that its extrapolation to the present time might not be faithful to the actual model under consideration. Likewise, the parameter values suggest very different behavior at $w(a=0)$ (the very distant past) than is seen in the actual model for all parameterizations considered. Therefore, the best fitting parameters could suggest unphysical behavior.  For example, extrapolating the CPL, EXP, or BA fits over all of cosmic history for the thawing quintessence model would lead one to conclude that the dark energy model it describes is phantom, which is clearly not the case as the actual model can never be phantom or cross the phantom line. In other words, it is important to not over-interpret any parameterization for dark energy as they can be misleading, even if they reproduce the observables of the dark energy models quite closely (more on this in Section \eqref{Sec:phenom}).

\subsection{Reproducing dark energy phenomenology}\label{Sec:phenom}

Up to this point, we have determined the best fitting $(w_0, w_a)$ values for all the parameterizations considered in this paper for very general models of thawing quintessence and non-minimal thawing quintessence. This will allow us, in Section \eqref{Sec:plane}, to compare their viability directly with the data (under these various different kinds of parameterizations/interpretations), as well as to develop intuition for features dark energy models must have to align with the data. 

However, it remains to be seen whether these various parameterizations for the equation of state $w(a)$ are suitable ways to go about describing the phenomenology of these microphysical dark energy models in the first place. After all, one look at Fig.~\eqref{fig:fits} and it is clear that none of these parameterizations have the ability to accurately describe the equation of state across cosmic history, and there is no guarantee that the physics of dark energy should be adequately captured by very simple two parameter functional forms like the ones considered here. However, what is most important to consider is not opinions on what functional forms dark energy should take; but rather, the ability for these parametrizations to accurately reproduce the phenomenology associated with the dark energy models they are intended to approximate.\footnote{For example, \cite{Garcia-Garcia:2019cvr} found that the CPL parameterization does not reproduce the observables for many  ``freezing'' quintessence models. There is no guarantee that these parameterizations are suitable and this is something that must be checked.} The ``best fits'' for $(w_0, w_a)$ in the previous section might be misleading if it turns out that they are actually a bad fit for the observables that they were fitted to. 

\begin{figure}[t]
    \centering
    \includegraphics[width=0.48\textwidth]{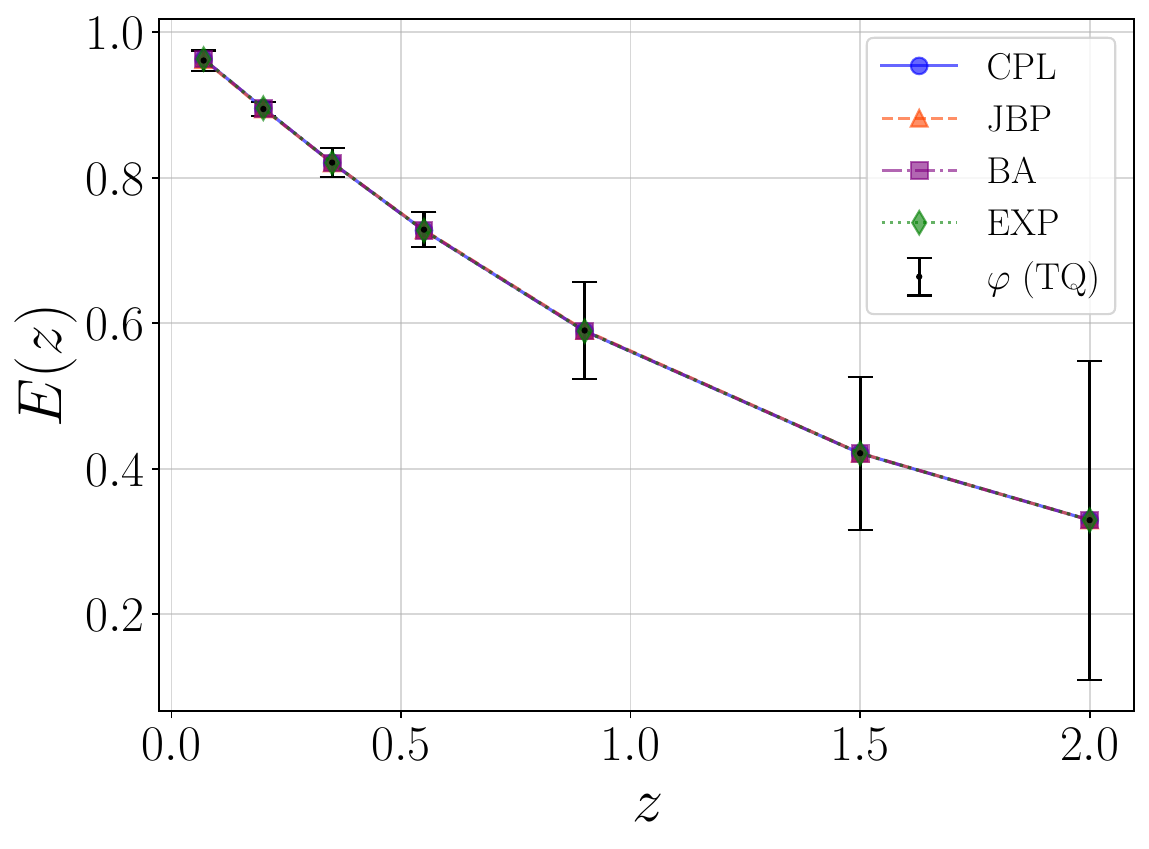}
    ~
    \includegraphics[width=0.48\textwidth]{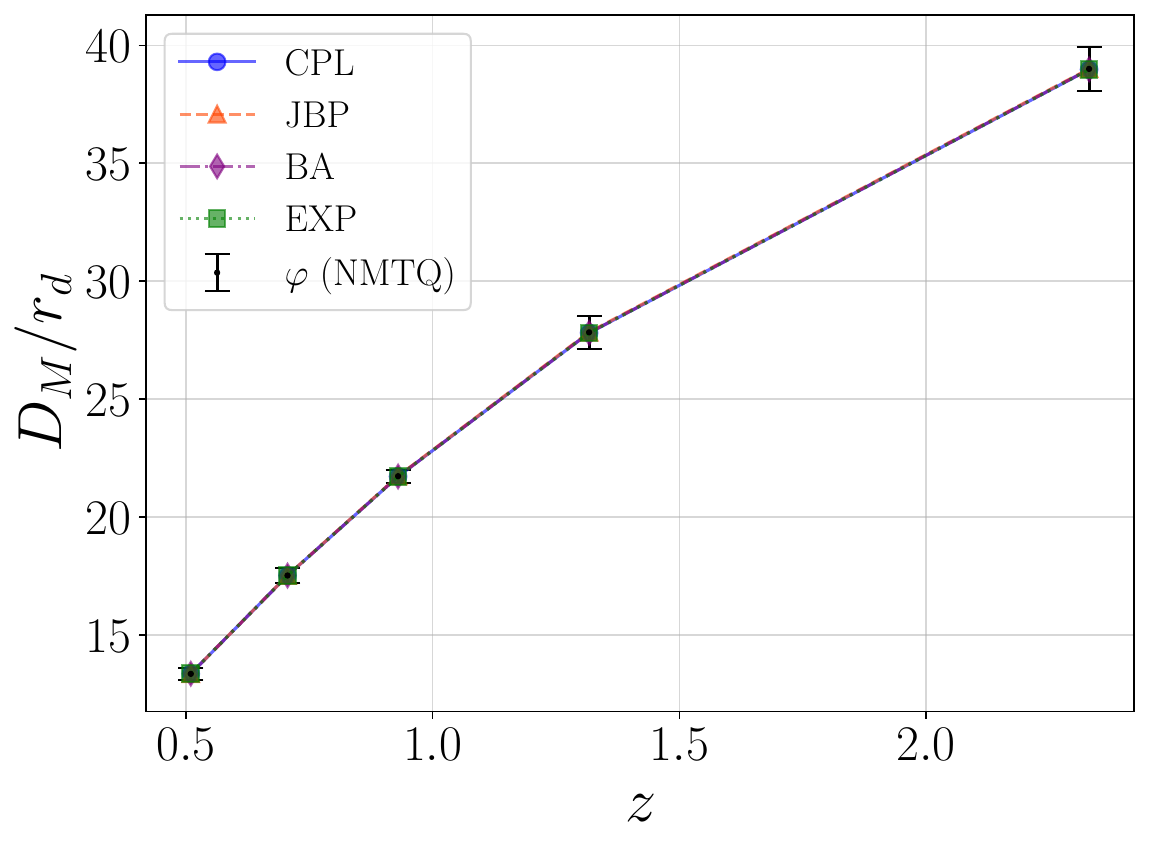}
    
    \caption{Comparison of some observables of dark energy theories with the observables produced by their best fit parameterizations. Left: SNe Ia observables $E(z)$ from the Pantheon+ compressed data for the thawing quintessence model given in Fig.~\eqref{fig:fits}. Right: DESI BAO observables $D_M / r_s$ for the non-minimal thawing quintessence model given in Fig.~\eqref{fig:fits} . As one can see, the best fit parameters for all parameterizations reproduce the predictions from the scalar field dark energy models to an excellent high degree of accuracy.} 
    \label{fig:example_observables}
\end{figure}

To investigate this, we take inspiration from e.g.~\cite{Traykova:2021hbr, Garcia-Garcia:2019cvr}, and we directly compare the predicted observables from the dark energy models with the $(w_0, w_a)$ observables determined from the best fit values. That is, we output predictions for quantities like $R$, $\ell_A$, $E(z)$, $D_M(z)$, etc (defined in Appendix \eqref{sec:compressed}) for the actual dark energy models and compare them to the values determined from the best fit $(w_0, w_a)$ values. As an example, in Fig.~\eqref{fig:example_observables} we show some of the exact observables that were used for the example models in Fig.~\eqref{fig:fits}: all parameterizations perform exceptionally well at reproducing the dark energy phenomenology for those specific models. However, these are just two dark energy models with specific choices of the microphysical model parameters. How do the dark energy parameterizations perform with respect to the more general class of theories we are considering? In order to answer this question, we examine the $\sim 15,000$ models used in the projection onto the $(w_0, w_a)$ plane (i.e. the wedges in Fig.~\eqref{fig:TQ}, in the next Section, generated as described in Section~\eqref{Sec:projection}). In particular, we look at those that utilized the Pantheon+ SNe Ia dataset.

\begin{figure}[t]
\centering
\includegraphics[width=.39\textwidth]{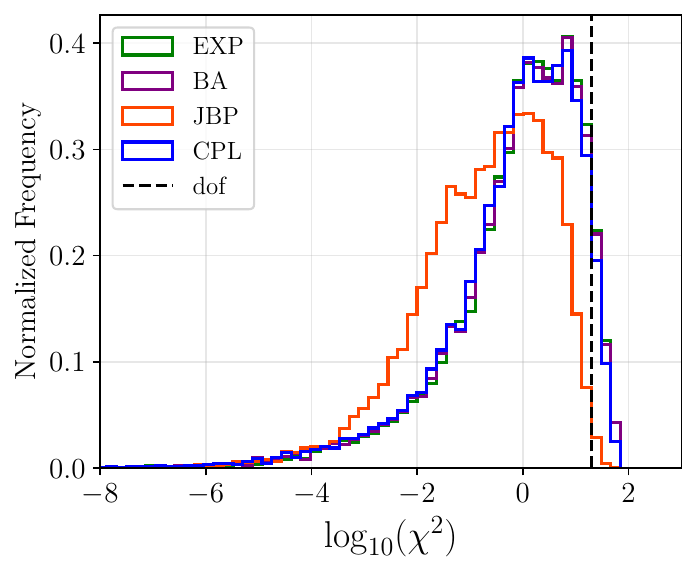}
~
\includegraphics[width=.39\textwidth]{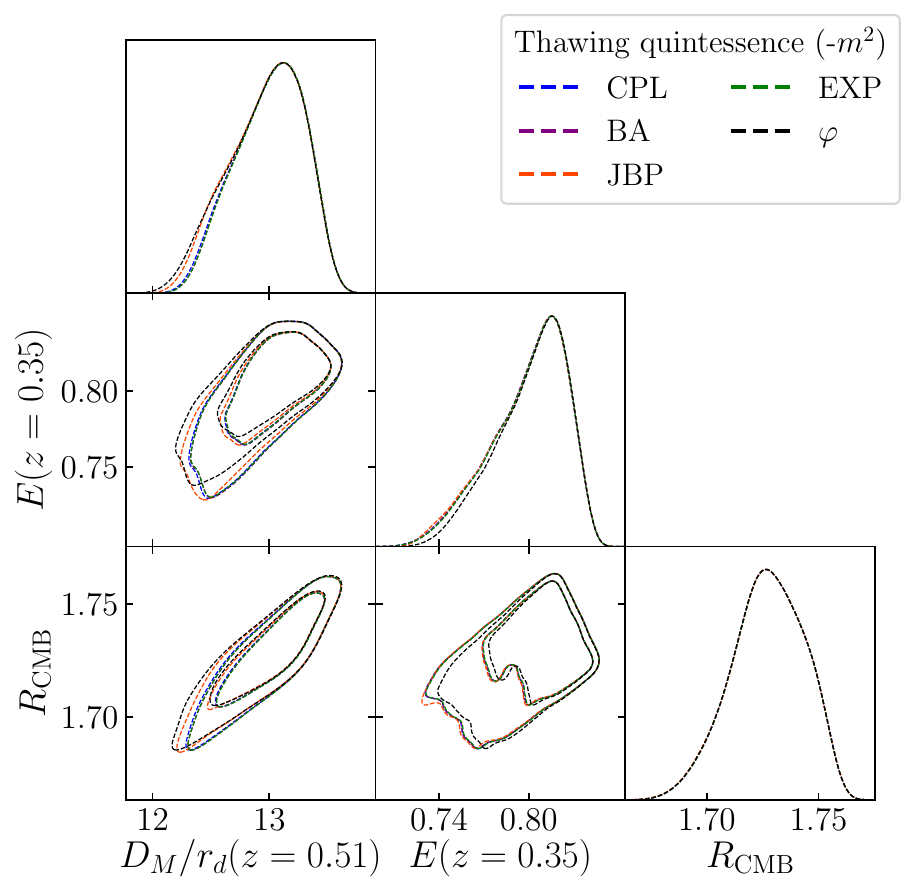}

\includegraphics[width=.39\textwidth]{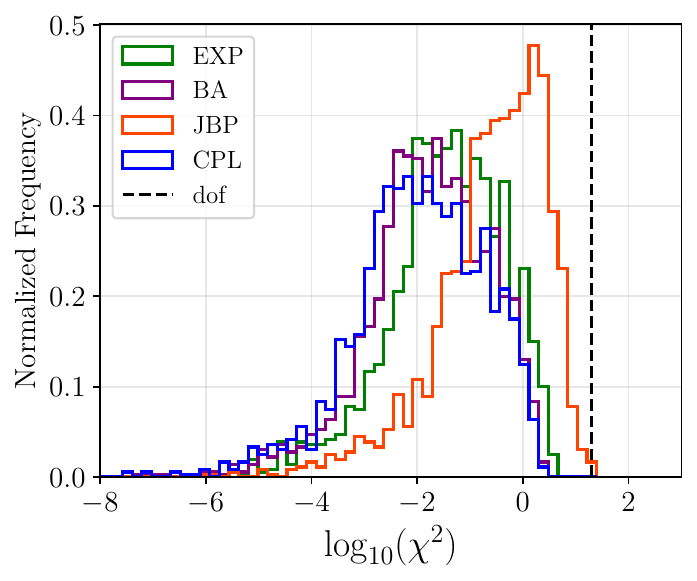}
~
\includegraphics[width=.39\textwidth]{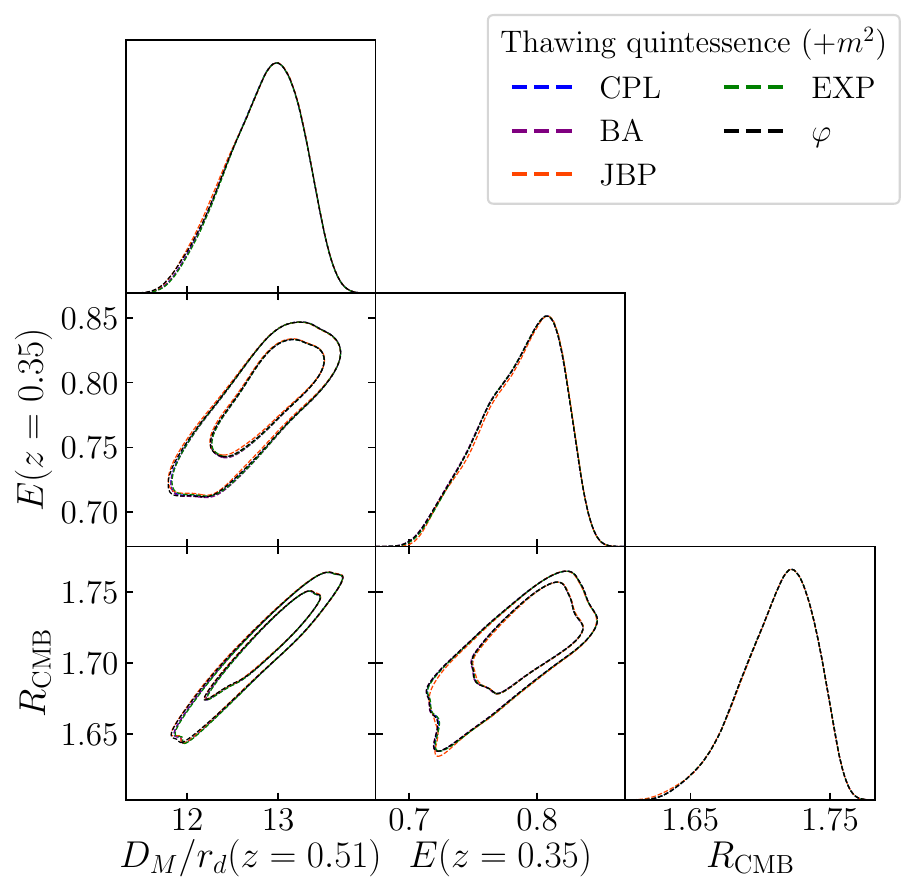}

\includegraphics[width=.39\textwidth]{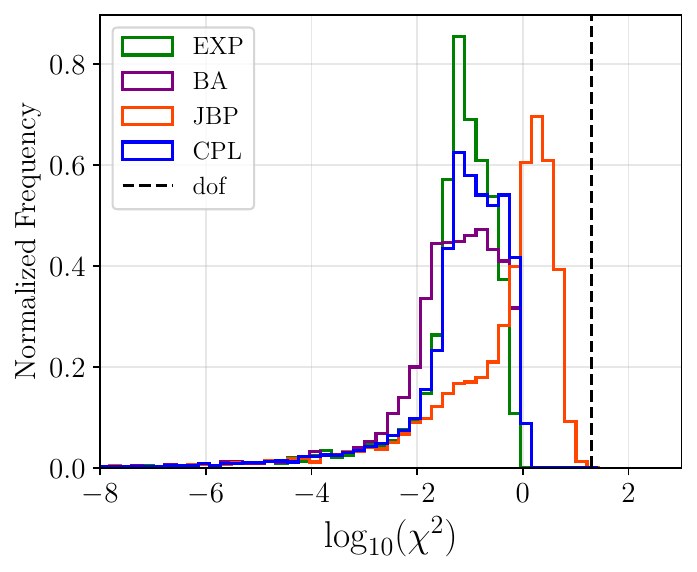}
~
\includegraphics[width=.39\textwidth]{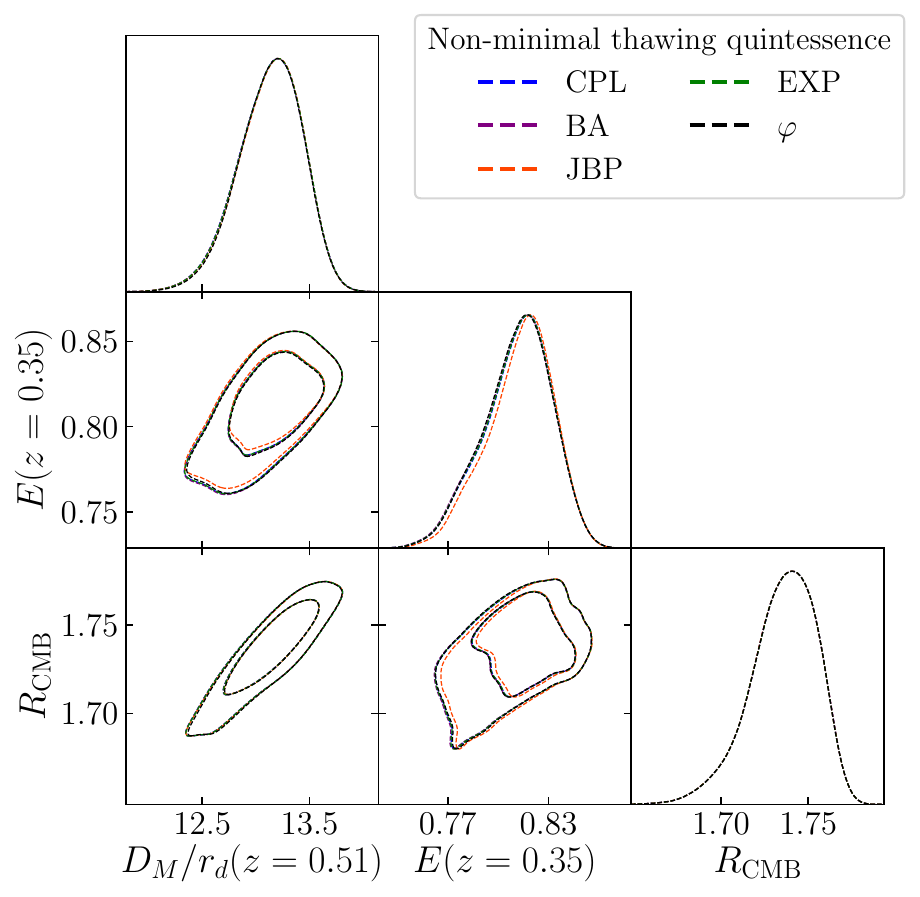}

\caption{Validation of the parametrizations to reproduce the phenomenology of the scalar field. From top to bottom: the hilltop branch of thawing quintessence, the positive mass branch of thawing quintessence, and non-minimal thawing quintessence. Left column: $\chi^2$ distribution from the parameterizations fit to the observables computed with the scalar field from Eq.~\eqref{chi2}. The black dashed line corresponds to the degrees of freedom ($\simeq 20$). Right column: 68\% and 95\% C.L. distributions of the fitted observables at different redshifts (dashed color lines), compared with the distribution obtained with the scalar field (dashed black line). All parametrizations accurately reproduce the theory predictions, but the JBP model does not perform quite as well as the others.\label{Fig:nmtq_hill_hist_phenom}}

\end{figure}

As can be seen in Fig.~\eqref{Fig:nmtq_hill_hist_phenom}, the fitting procedure we have used reproduces the predicted cosmological observables of both thawing quintessence and non-minimal thawing quintessence very accurately for \textit{all} parameterizations considered. While there are some differences between the parameterizations, given that we have $\text{DoF} \simeq 20$ here, when examining the $\chi^2$ determined from minimizing Eq.~\eqref{chi2} we consistently have that $\chi^2/\text{DoF} < 1$ or $\simeq 1$ across the board regardless of the dark energy model or parameterization considered. For both the positive branch of thawing quintessence and the non-minimal thawing quintessence model, $\sim
99.99\%$ of the models for all parameterizations have $\chi^2/\text{DoF} < 1$. For the hilltop branch it is a bit more interesting as the JBP parameterization is such that $\sim
99\%$ of the models have $\chi^2/\text{DoF} < 1$, whereas for the other parameterizations it is $\sim
95\%$ of the models that have $\chi^2/\text{DoF} < 1$. 

It is interesting to note that while the JBP parametrization outperforms the others in the case of the hilltop branch of thawing quintessence, it systematically fits worse than the other parameterizations for the positive branch of thawing quintessence and non-minimal thawing quintessence (although as noted above it still performs very well). Given that the JBP parameterization by construction is essentially agnostic about $w(a)$ at $a=0$ and gives significant weight towards more recent evolution, it is perhaps not surprising that the hilltop branch of thawing quintessence, which can evolve very rapidly and non-linearly at more recent times, is slightly better represented by the JBP model which also evolves non-linearly with the scale factor.

Regardless, this will have a small impact on the results shown in next Section. However, in general, for all intents and purposes, it does not matter which parametrization one uses. All of them are capable of reproducing the phenomenology of the most straightforward and widely explored classes of dynamical dark energy theories (thawing quintessence with and without a non-minimal coupling) to a high degree of accuracy, and all of them provide, as we will show in the next Section, similar insights regarding the viability or non-viability of various dark energy models when one compares theory predictions and actual data measurements within the parameterizations. In other words, our conclusions regarding the dynamical dark energy models are robust to our choices for how we parameterize the equation of state.

\subsection{Data constraints and viability of the dark energy models}\label{Sec:plane}

\begin{figure}[t]
  \centering
  \includegraphics[width=0.80\columnwidth]{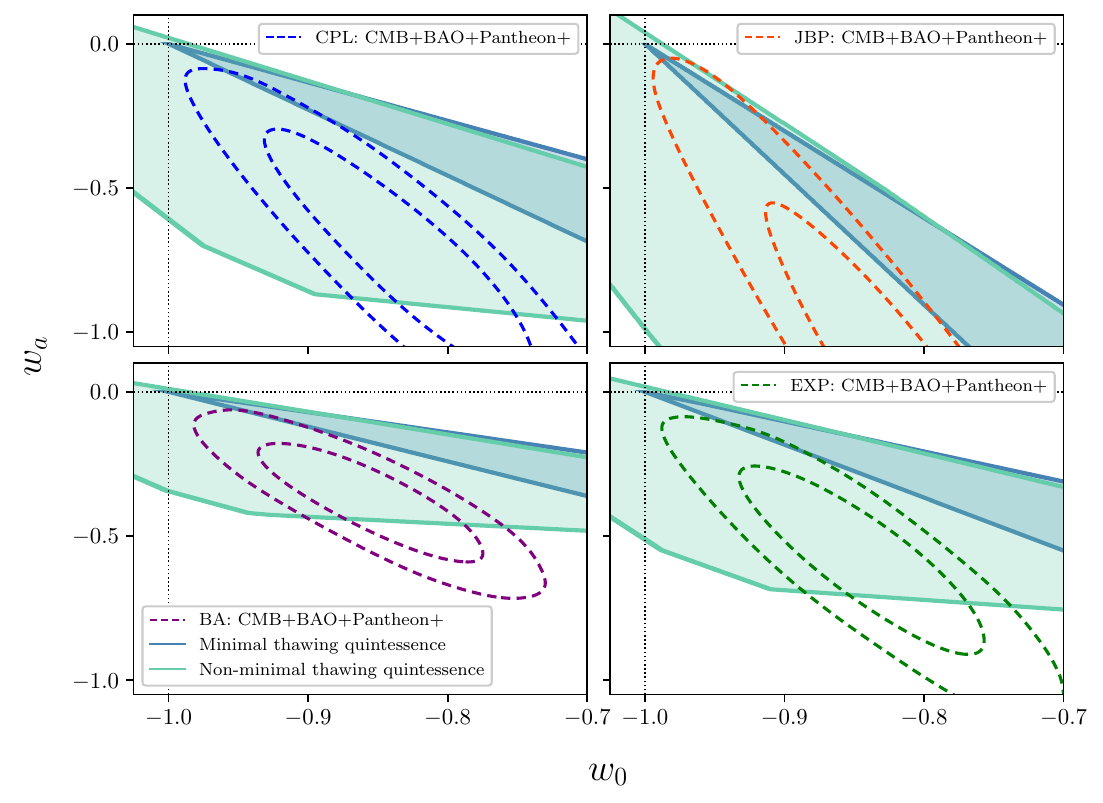}
  
  \includegraphics[width=0.80\columnwidth]{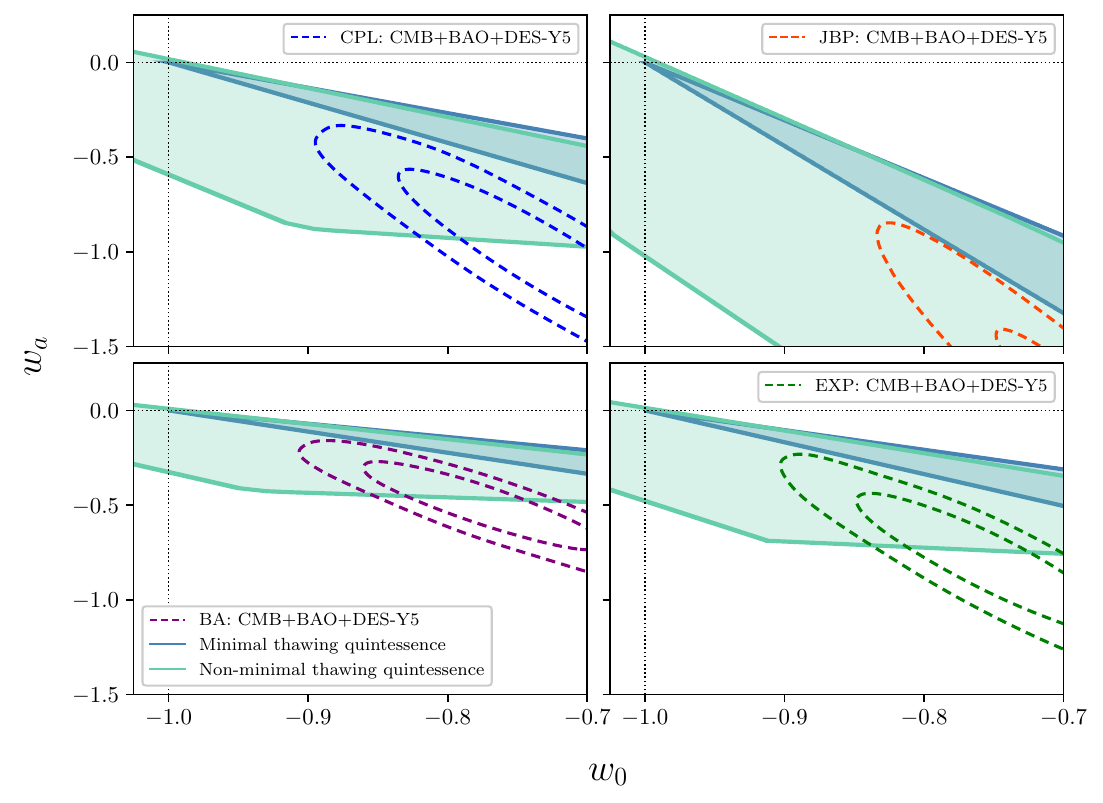}
  
  \caption{68\% and 95\% C.L. posterior distribution of the various parameterizations (dashed lines) from the combined BAO+CMB+SNe data, where we use the data sets coming from  DESI \cite{DESI:2024kob}, Pantheon+ (top) \cite{Scolnic:2021amr},  DES-Y5 (bottom) \cite{DES:2024tys}, and Planck+ACT lensing \cite{Planck:2018vyg, Planck:2019nip,ACT:2023kun, ACT:2023dou}. Overlayed are the predictions for thawing quintessence and non-minimal thawing quintessence in terms of their $(w_0, w_a)$  parameters, obtained by fitting Eq.~\eqref{chi2}. Both minimal and non-minimal thawing quintessence behave similarly regardless of parameterization is used. While it is not visible on this plot as all parameterizations share the same axes here, the lower bound of the non-minimal model cuts off in the upper right JBP plot in a similar location as it does with all the other parameterizations.}
  \label{fig:TQ}
\end{figure}

Fig.~\eqref{fig:TQ} shows the data constraints from BAO, CMB and SNe Ia data, and the corresponding theory projections for each parametrizations. The top Figure shows the results when using Pantheon+, whereas the lower one when DES-Y5 SNe Ia are used. We note that all data constraints agree with \cite{Giare:2024gpk} and display a robust preference for dynamical dark energy, more than $2\sigma$ away from $(w_0, w_a) = (-1, 0)$. In the same plots we can see the projection of minimally and non-minimally coupled thawing quintessence models onto the $(w_0, w_a)$ plane. As we expected from the results in the previous Section~\eqref{Sec:phenom}, we find that all parameterizations produce very similar results. 

Focusing first on thawing quintessence, we see that it sweeps out a wedge in the $(w_0, w_a)$ plane whose inclination follows that from the data contours. As discussed in \cite{Wolf:2023uno, Wolf:2024eph}, this wedge is constrained by the fact that, even though $w(a)$ can in principle evolve very sharply with these models as $|-m^2|$ is made larger, the larger $m^2$ gets the more this evolution is weighted towards very recent redshifts, which means it is outside the range of redshifts that most of the data is probing. So, fitting $(w_0, w_a)$ over the redshifts that are actually probed will not capture extremely late time evolution. Regardless of which parameterization is used, the thawing quintessence wedges and the data contours are almost entirely disjoint, regardless of having used Pantheon+ or DES-Y5 SNe Ia. Even when DES-Y5 moves the data constraints further away from $\Lambda$CDM, the overlap with the theory contours remain minimal, meaning that data is still not favoring thawing quintessence. 

This is confirmed when we compare the various parametrizations and thawing quintessence in terms of their ability to fit the complete (uncompressed) data. To do so, we find the best fitting model of thawing quintessence (i.e. that maximizes the likelihood ${\cal L}$ or, alternatively, minimizes $\chi^2=-2\ln {\hat{\cal L}}$ for the given datasets\footnote{Here, we use the \textit{lite} version of Planck likelihood to do the minimization as it has substantially fewer parameters than the full likelihood and is thus more efficient.}).  We do the same with $\Lambda$CDM and with the CPL, JBP, EXP, and BA parameterizations. The best-fit $\chi^2$ can be seen in Table \eqref{tab:chi2_dark_energy} and show that both thawing quintessence and all parametrizations can fit the data better than $\Lambda$CDM, in agreement with \cite{Giare:2024gpk}. In particular, while the scalar field improves the $\chi^2$ of $\Lambda$CDM only by $\Delta \chi^2 = -2.9$, the parametrizations obtain $\Delta \chi^2 = -6.7,\, -6.9,\, -7.1$, for CPL, EXP and BA. The worst performing parametrization is JBP, with $\Delta \chi^2 = -4.9$, and this worse fit is the reason why in Fig.~\eqref{fig:TQ}, for JBP, the wedge of the theory projection has a bigger overlap with the data constraints: both JBP and thawing quintessence are relatively close in terms of their ability to fit the expansion history data, meaning that when thawing quintessence is projected into JBP parameter space there will be more overlap than with the other parameterizations which are further away from thawing quintessence in terms of their ability to fit the data.

\begin{table}
\centering
\begin{tabular}{lcccc}
\toprule
DE Model & BAO+CMB+Pantheon+ & BAO+CMB+DES-Y5 \\
\midrule
$\Lambda$CDM        &  2436.8 (ref)    &     2680.3 (ref)        \\
CPL                 &  2430.1 (-6.7)  &     2663.8 (-16.5) \\
EXP                 &  2429.9 (-6.9)  &     2663.9 (-16.4) \\
BA                  &  2429.7 (-7.1)  &     2664.1 (-16.2) \\
JBP                 &  2431.9 (-4.9)  &     2664.4 (-15.9) \\
TQ                  &  2433.9 (-2.9)  &     2668.2 (-12.1) \\
NMTQ                &  2424.0 (-12.8) &     2659.8 (-20.5) \\
\bottomrule
\end{tabular}
\caption{Best fit $\chi^2$ values for $\Lambda$CDM and the various dark energy parameterizations and models considered in this paper. In parenthesis, the difference with respect to $\Lambda$CDM best fit $\chi^2$. These values were obtained by minimizing $\chi^2$. }\label{tab:chi2_dark_energy}
\end{table}

Moving onto non-minimal thawing quintessence, we again find that all parameterizations produce very similar results. However, this time the dark energy models are not confined to a narrow wedge, but rather sweep out across the $(w_0, w_a)$ plane for all parametrizations, including covering much of the favored regions picked out by the data as can be seen in Fig.~\eqref{fig:TQ}. This suggests that this model provides at the very least a comparable fit to the data as the various parameterizations. As discussed in \cite{Wolf:2024stt}, the ability for these models to occupy this region of the $(w_0, w_a)$ parameter space is due to their phantom crossing behavior. They have the dynamical freedom to be phantom throughout much of their history, before sharply thawing to cross the phantom divide. However, this thawing can occur within the redshifts where cosmological probes have robust coverage (see the example models depicted in Fig.~\eqref{fig:fits} here or Fig.~(1) in \cite{Wolf:2024stt}). Consequently, the fitted $(w_0, w_a)$ parameters for this kind of dark energy evolution will map into the lower left parts of the $(w_0, w_a)$ plane as this region implies that dark energy is thawing and has undergone significant temporal evolution.

Regardless of which parameterization is used, the non-minimal thawing quintessence wedges share significant overlap with the contours determined from the data. What happens when we compare how this model fits the complete data to how the various parameterizations perform? Minimizing $\chi^2$ for the non-minimal thawing quintessence models validates this intuition, finding not only that this dark energy model offers comparable fits to the various parameterizations, but also outperforms them. As can be seen in Table \eqref{tab:chi2_dark_energy}, this model consistently offers the best fit to the data of all the models considered here independently of the SNe sample we use (i.e.~Pantheon+ or DES-Y5 SNe). Moreover, it is able to improve $\Lambda$CDM fit by $\Delta \chi^2 = -20.5$ when using the DES-Y5 sample\footnote{$\Delta \chi^2 \simeq -12$ was found for a similar non-minimal dark energy model in \cite{Wolf:2024stt} using the Pantheon+ dataset as is the case here.} where the best fit parameters for the non-minimal model are $(\xi, V_0, \beta, m^2)\simeq(2.9, 0.55, 5.4, -5.6)$. 

There are, however, a couple of points worth briefly commenting on. Naively, if we use the  Akaike Information Criterion
(AIC), which is given by
$
 {\rm AIC}=2k+\chi^2    
$,
where $k$ is the number of parameters in the model, we find that $\Delta {\rm AIC}\simeq 12.5$ because the non-minimal model has 5 parameters (the 4 mentioned above plus one additional parameter $\dot{\phi}_{ini}$ for the initial conditions) while $\Lambda$CDM has only one ($\Lambda$). Even accounting for the additional
number of parameters, this is still a notable improvement that, as mentioned above, merits
further investigation. See \cite{Wolf:2025jed} for further discussion and statistical analysis.

Another point to consider is that, while Section \eqref{Sec:phenom} demonstrated that the various dark energy parameterizations can all accurately reproduce the phenomenology of both minimal and non-minimal thawing quintessence, the fact that the non-minimal model can achieve a lower $\chi^2$ than any of the parameterizations does show that there is \textit{some} information that is being lost when the full dark energy model is projected into the parameterizations' parameter spaces. However, this makes sense as fitting directly to the observables using Eq.~\eqref{chi2} was only fit to \textit{background} observables that represent compressed versions of the full data (with the exception of the BAO data which requires no compression). The biggest difference between the compressed (background) observables and the full data will be in the CMB power spectra and CMB lensing data from Planck and ACT, where one must compute the full linear cosmological perturbations to compare with the data. And indeed, while we find that the non-minimal model reproduces similar $\chi^2$ with respect to the SNe and BAO components of the full data when compared with the various parameterizations, the improvement in overall $\chi^2$ ($\Delta \chi^2 \simeq 4$ in the case of the DES-Y5 dataset) over those of the various parameterizations was largely driven by the Planck and ACT components of the data. So while these parameterizations clearly perform exceptionally well at reproducing dark energy phenomenology at the background level, it is perhaps not surprising that there is some information lost at the perturbative level.

Of course, the actual comparison to the full data is the most important in assessing the viability of the model. However, converting a dark energy model's observables into $(w_0, w_a)$ parameters is useful as it gives us a reliable look at the viability of the model with respect to the data, as well as further perspective on what features a model needs in order to overlap with the data, just as comparing inflationary models' predictions for $(r, n_s)$ to the data constraints does for inflation. While we caution against overinterpretting any specific parameterization due to the issues highlighted here and elsewhere, it is nonetheless the case that they can be informative. Here, all of these parameterizations suggest that the dark energy equation of state $w(a)$ is thawing ($dw/da > 0$), has undergone sharp temporal evolution (seen in the relatively large magnitude of $\vert -w_a \vert$), and has possibly undergone a recent phantom crossing (the combination of $w_0 > -1$ and large $\vert -w_a \vert$). This suggests that a model of the type constructed in Eq.~\eqref{eq:NMTQ} will be a good fit to the data, and indeed this model is found to outperform both $\Lambda$CDM and thawing quintessence (given by Eq.~\eqref{eq:TQ}) as it occupies the exact region suggested by the data (interpreted in terms of the various parameterizations considered here).

\section{Discussion}\label{Sec:conclude}

The recent DESI results, in conjunction with other cosmological data, paints an intriguing picture regarding the fundamental nature of dark energy. However, as we have seen, the process of interpretting the data involves making what are, to a large extent, arbitrary choices for the parameterization of the dark energy equation of state. In this paper, we explored the robustness of conclusions about dynamical dark energy models—both minimally and non-minimally coupled scalar field models—across four widely used two-parameter models: the Chevallier-Polarski-Linder (CPL), Jassal-Bagla-Padmanabha (JBP), Barboza-Alcaniz (BA), and exponential (EXP) parameterizations \cite{Chevallier:2000qy,Linder:2002et,Pan:2019brc, Jassal:2005qc, Barboza:2008rh}.

Our analysis reveals that assessing the viability of these dynamical dark energy models is remarkably robust to the choice of parameterization. By mapping the dynamical dark energy models into the $(w_0, w_a)$ parameter space for each parameterization, we demonstrated both that our conclusions regarding their viability (or how well they describe the data) are not overly subject to particular choices of parameterization and also that all of these various parameterizations effectively capture the phenomenology of the full dark energy models. This demonstrates that, at least for these very broad and simple kinds of microphysical dark energy models considered here, any of these choices of parameterization are sufficient to interpret the data and make assessments regarding the viability of such dark energy proposals.

\acknowledgments
We are very grateful to William Giar\`{e} for helpful conversations.

We acknowledge the \texttt{hi\_class} developers, Emilio Bellini, Miguel Zumalac\'arregui and Ignacy Sawicki, for sharing a private version of the code.
WJW is supported by the HAPP Centre at St.~Cross College, University of Oxford. CGG is supported by the Beecroft Trust.
PGF is supported by STFC and the Beecroft Trust. For the purposes of open access, the authors have applied a Creative Commons Attribution (CC BY) licence to any Author Accepted Manuscript version arising.

{\it Software}:  We made extensive use of {\tt hi\_class} \cite{hi_class_1,hi_class_2,Blas:2011rf}, {\tt Cobaya} \cite{Torrado:2020dgo} and the {\tt numpy} \cite{oliphant2006guide, van2011numpy}, {\tt scipy} \cite{2020SciPy-NMeth}, and {\tt matplotlib} \cite{Hunter:2007} python packages. 

\appendix
\section{Compressed data for theory priors}\label{sec:compressed}
Here we describe the fitting procedure that can be used to compute $(w_0, w_a)$ predictions for the dark energy models described in Section \eqref{Sec:DEmodels}. When we adopt any parameterization for the equation of state of dark energy $w(a)$, we are essentially choosing a form of data compression, which then allows us to see where the data itself lies in terms of the 68\% and 95\% C.L. regions of the parameters we have adopted. However, while adopting certain parameterizations for dark energy may be very useful and even necessary in some circumstances, dark energy itself is not exactly or fundamentally described by a particular parameterization, but rather, by specific microphysical models such as, for example, the ones highlighted in 
Section \eqref{Sec:DEmodels}. Thus, in assessing the viability of various dark energy proposals, it is helpful to understand where these microphysical models lie (or rather, what they would predict) in terms of the parameters that have been adopted in the analysis of the data. 

Adopting a particular parametrization allows us, in a simplified way which can be easily visualized, to compare what the data is preferring with where the microphysical model would naturally lie. Put another way (and ignoring other cosmological parameters for now), we can compare the likelihood for the parameters $(w_0,w_a)$ dictated by the data, $D$, i.e.~${\cal L}(w_0,w_a| D)$, with the priors dictated by a microphysical model with parameters ${\vec \alpha}$, i.e.~${\cal P}_r(w_0,w_a|{\vec \alpha})$. While, ultimately, for any microphysical model, we are interested in the posterior distribution on ${\vec \alpha}$, given by
\begin{eqnarray}
{\cal P}({\vec \alpha})=\int dw_0dw_a {\cal P}_r(w_0,w_a|{\vec \alpha}){\cal L}(w_0,w_a| D), \nonumber
\end{eqnarray}
by separating out ${\cal L}$ and ${\cal P}_r$ we are able to see if they are overlapping, or disjoint, i.e., if the microphysical model is a ``good'' or ``bad'' one, given the data. In other words, adopting a particular parameterization will allow us to determine a likelihood for the data itself in terms of those parameters, and we would then like to know whether a dark energy model's predictions for what we would observe in terms of those parameters (i.e.\ its prior distribution in that parameter space) shares overlap with what the data itself selects for these parameters.

This is exactly analogous to what is done in the study of inflation models. We make observations that constrain the tensor-to-scalar ratio $r$ and the scalar spectral index $n_s$, and then use these constraints to determine the observationaly favored region of the $(r, n_s)$ plane. To evaluate whether or not specific inflationary models are viable, we compute their predictions for $(r, n_s)$ and compare these to those constraints (e.g.~\cite{Planck:2018jri, Kallosh:2019jnl, Martin:2013tda, Wolf:2024lbf, Dodelson:1997hr}). However, unlike with inflation, as we have seen there are a number of distinct parameterizations for dark energy and these parameters are a bit further from the actual observables. So, in order to actually determine how well a dark energy model maps onto these constraints, we must determine what a dark energy model would predict for $(w_0, w_a)$ by using a procedure that mimics what was done for determining where the data itself lies in terms of the parameters. In the dark energy literature, it is common to generate a dark energy model's equation of state and ``fit'' this equation of state over some recent range of redshifts using one of the parameterizations directly as a proxy for determining where a model might lie on the $(w_0, w_a)$ plane. However, as discussed in \cite{Wolf:2023uno, Wolf:2024stt, Wolf:2024eph}, because we do not measure the equation of state directly, doing this can often lead to highly misleading results. In other words, we must generate the actual observables for a given dark energy model, and then determine what $(w_0, w_a)$ parameters best reproduce those observables given the known errors and uncertainties of the data. This will allow us to make an ``apples to apples'' comparison.

In \cite{Wolf:2024eph}, we develop such a procedure. Essentially, one can take the data discussed above and ``compress'' them into a set of ``measured'' physical parameters and their covariance. The CMB data can be compressed into an acoustic scale $\ell_{\mathrm{A}}$ and shift parameter $R\left(z_{*}\right)$ at the photon decoupling redshift $z_{*}$ \cite{Chen:2018dbv}:
\begin{equation}\label{eq:cmb1}
\ell_{\mathrm{A}}=\left(1+z_*\right) \frac{\pi D_{\mathrm{M}}\left(z_*\right)}{r_s\left(z_*\right)},
\end{equation}
\begin{equation}\label{eq:cmb2}
R\left(z_*\right) \equiv \frac{\left(1+z_*\right) D_{\mathrm{M}}\left(z_*\right) \sqrt{\Omega_m H_0^2}}{c},
\end{equation}
where $r_s$ is the co-moving sound horizon and $D_M$ is the angular diameter distance.

The SNe data are directly sensitive to the luminosity distance:
\begin{equation}\label{eq:D_L}
    D_L = \frac{1+z}{H_0} \int_0^z \frac{\md z'}{E(z')},
\end{equation}
and can be compressed into measurements of $E(z) \equiv H(z) / H_0$ at different redshifts \cite{Riess:2017lxs}. Here, we use a cubic spline $1/E(z) = {\rm spline}(1/E(z_i))$, with nodes $1/E(z_i)$ defined at $z = \{0.07, 0.2, 0.35, 0.55, 0.9, 1.5, 2, 2.5\}$.

The BAO data are sensitive to the angular diameter distance or Hubble parameter, and consequently can be used without any further compression. As mentioned, the quantities measured are $D_M$, $D_H = c/H$, and $D_V = (z D_M^2 D_H)^3$ \cite{DESI:2024mwx},
with the angular diameter distance defined as:
\begin{equation}\label{eq:D_M}
D_{\mathrm{M}}(z)=\frac{c}{H_0 \sqrt{\Omega_{\mathrm{K}}}} \sinh \left[\sqrt{\Omega_{\mathrm{K}}} \int_0^z \frac{d z^{\prime}}{H\left(z^{\prime}\right) / H_0}\right] .
\end{equation}
These measurements also depend on the so-called ``drag'' epoch where baryons decoupled from photons, defined as,
\begin{equation}
r_{\mathrm{d}}=\int_{z_{\mathrm{d}}}^{\infty} \frac{c_{\mathrm{s}}(z)}{H(z)} d z.
\end{equation}
Thus BAO measurements constrain the ratios $D_M(z)/r_{\mathrm{d}}$, $D_H(z)/r_{\mathrm{d}}$, and $D_V(z)/r_{\mathrm{d}}$.

\begin{figure}[t]
  \centering
  
  \subfigure[]{
      \includegraphics[width=0.45\textwidth]{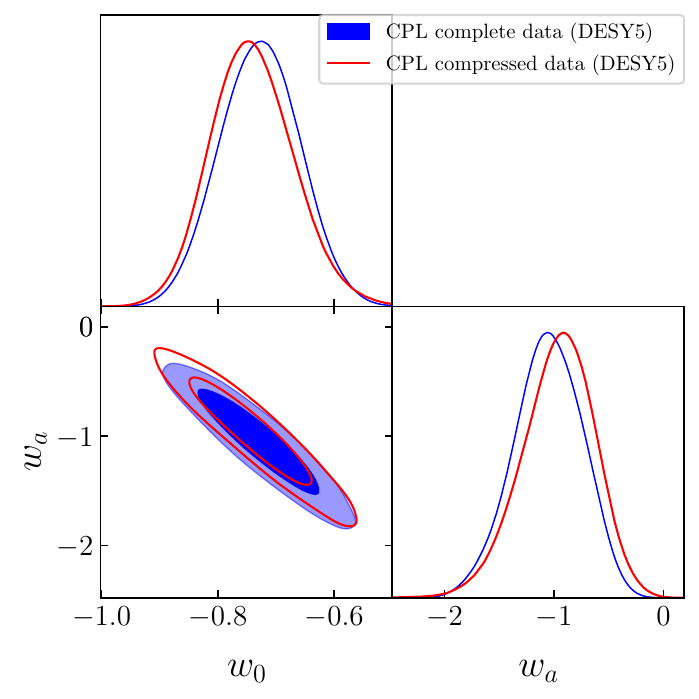}
  }
  \qquad
  \subfigure[]{
      \includegraphics[width=0.45\textwidth]{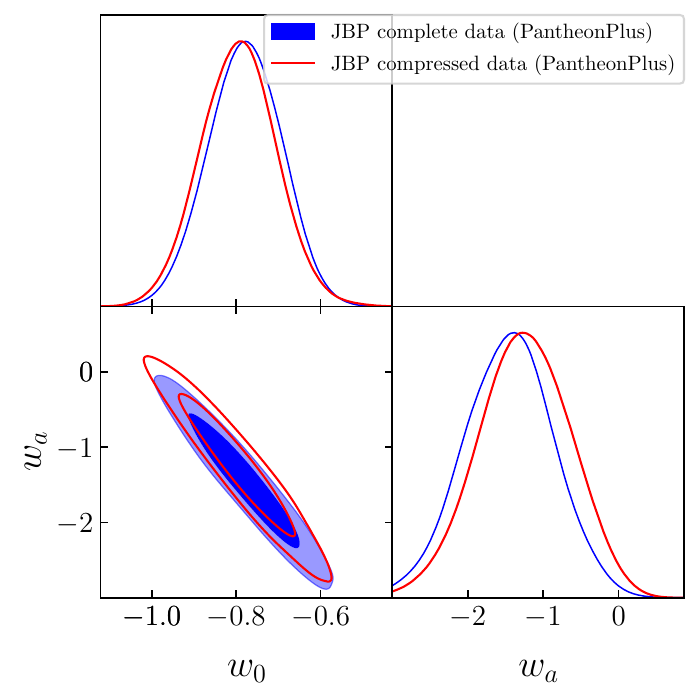}
  }
  
  \subfigure[]{
      \includegraphics[width=0.45\textwidth]{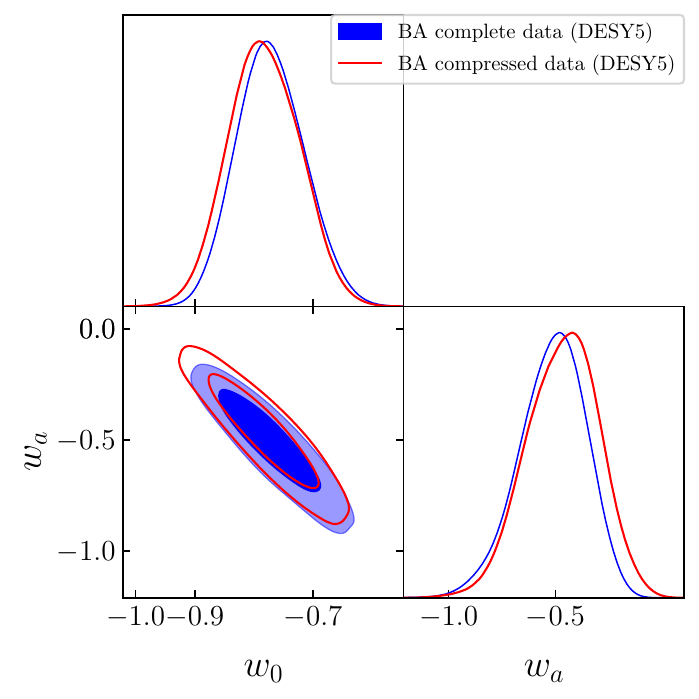}
  }
  \qquad
  \subfigure[]{
      \includegraphics[width=0.45\textwidth]{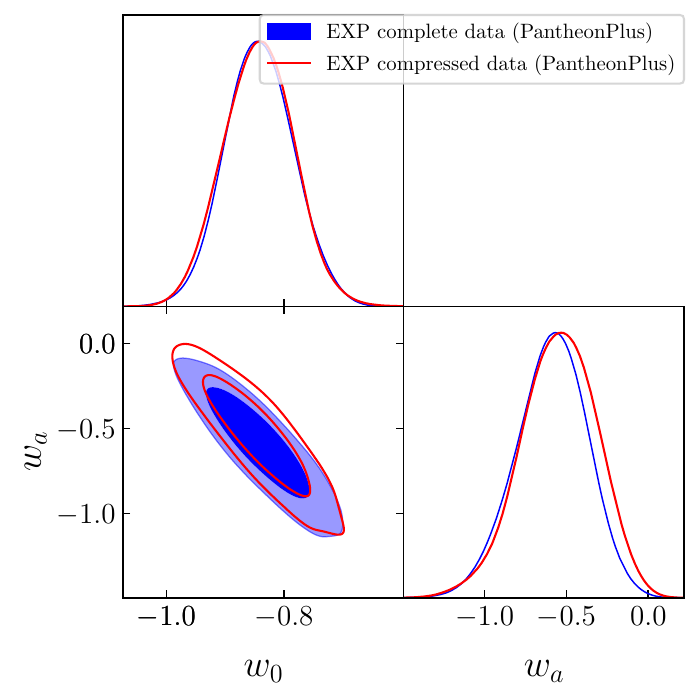}
  }
  
  \caption{Data compression validation. We show the 68\% and 95\% C.L. posterior distribution of the dark energy parameters from the full combined BAO+CMB+SNe data \cite{DESI:2024kob, Scolnic:2021amr, DES:2024tys, Planck:2018vyg, Planck:2019nip,ACT:2023kun, ACT:2023dou}, compared with the compressed data for particular choices of dark energy parameterization and datasets. Here, the compressed data refers to the compressed CMB data and the compressed SNe Ia data (where we have denoted which SNe data was used for each specific plot in parenthesis in each legend). As the DESI BAO data does not need to be compressed for the fitting procedure, the standard full DESI data is used in all of the above. As is clear from the results, the compressed data reproduces the results of the full data to a high degree of accuracy, so we can be sure that we are not losing much information when determining the best fit $(w_0, w_a)$ parameters for a given dark energy model when using the compressed data. \label{fig:compressed_v_actual}}
\end{figure}

In order to validate the compression,  
we check that the compressed data reproduces the posterior distributions of the full datasets. Fig.~\eqref{fig:compressed_v_actual} demonstrates that this is the case by considering a few examples (one from each of the four parameterizations) to show that the compressed data reproduces the results of the full data to a high level of accuracy. One can also see the means, errors, and correlation matrices of the SNe and CMB compressed variables in Table \eqref{tab:SNe_DES_CMB_compressed}.

With this in hand, we are now in position to convert the phenomenological behavior of dark energy models into $(w_0, w_a)$ parameters in a way that directly uses the data (rather than simply fitting the equation of state as a proxy). That is, we generate the relevant observables discussed above for a dark energy model (replacing the actual data measurements with these predictions from the dark energy model) and then determine the best fit $(w_0, w_a)$ parameters \textit{using the associated errors and covariances of the actual data} by minimizing Eq.~\eqref{chi2}.

\begin{table}[t]
    \centering
    \begin{tabular}{|c|c|ccccccc|}
    \hline
    \multicolumn{9}{|c|}{Pantheon+ SNe Compressed Likelihood} \\
    \hline
    Redshift & Mean $1/E(z)$ $\pm$ error & \multicolumn{7}{c|}{Correlation Matrix} \\
    \hline
    0.07 & $0.947 \pm 0.014$ & 1.00 & 0.15 & 0.59 & 0.10 & 0.15 & -0.05 & 0.07 \\
    \hline
    0.20 & $0.8982 \pm 0.0095$ & 0.15 & 1.00 & -0.16 & 0.20 & 0.04 & 0.04 & 0.03 \\
    \hline
    0.35 & $0.815 \pm 0.020$ & 0.59 & -0.16 & 1.00 & -0.24 & 0.21 & -0.14 & 0.12 \\
    \hline
    0.55 & $0.664 \pm 0.024$ & 0.10 & 0.20 & -0.24 & 1.00 & -0.50 & 0.32 & -0.11 \\
    \hline
    0.90 & $0.790 \pm 0.066$ & 0.15 & 0.04 & 0.21 & -0.50 & 1.00 & -0.66 & 0.25 \\
    \hline
    1.50 & $0.18 \pm 0.11$ & -0.05 & 0.04 & -0.14 & 0.32 & -0.66 & 1.00 & -0.34 \\
    \hline
    2.00 & $0.46 \pm 0.22$ & 0.07 & 0.03 & 0.12 & -0.11 & 0.25 & -0.34 & 1.00 \\
    \hline
    \end{tabular}

    \vspace{10pt} 

    \begin{tabular}{|c|c|ccccccc|}
    \hline
    \multicolumn{9}{|c|}{DES-Y5 SNe Compressed Likelihood} \\
    \hline
    Redshift & Mean $1/E(z)$ $\pm$ error & \multicolumn{7}{c|}{Correlation Matrix} \\
    \hline
    0.07 & $0.9183 \pm 0.0238$ & 1.00 & 0.26 & 0.66 & 0.37 & 0.27 & -0.10 & 0.01 \\
    \hline
    0.20 & $0.8743 \pm 0.0106$ & 0.26 & 1.00 & -0.06 & 0.30 & -0.02 & 0.09 & -0.03 \\
    \hline
    0.35 & $0.7753 \pm 0.0212$ & 0.66 & -0.06 & 1.00 & -0.11 & 0.47 & -0.27 & 0.05 \\
    \hline
    0.55 & $0.7355 \pm 0.0178$ & 0.37 & 0.30 & -0.11 & 1.00 & -0.30 & 0.29 & -0.06 \\
    \hline
    0.90 & $0.4677 \pm 0.0379$ & 0.27 & -0.02 & 0.47 & -0.30 & 1.00 & 0.04 & 0.01 \\
    \hline
    1.50 & $0.5721 \pm 0.4165$ & -0.10 & 0.09 & -0.27 & 0.29 & 0.04 & 1.00 & -0.06 \\
    \hline
    2.00 & $1.1067 \pm 0.5769$ & 0.01 & -0.03 & 0.05 & -0.06 & 0.01 & -0.06 & 1.00 \\
    \hline
    \end{tabular}

    \vspace{10pt} 

    \begin{tabular}{|c|c|cccc|}
    \hline
    \multicolumn{6}{|c|}{CMB Compressed Likelihood} \\
    \hline
    Variable & Mean $\pm$ error & \multicolumn{4}{c|}{Correlation Matrix} \\
    \hline
    $R$ & $1.7508 \pm 0.0041$ & 1.00 & 0.35 & -0.64 & -0.73 \\
    \hline
    $l_a$ & $301.544 \pm 0.084$ & 0.35 & 1.00 & -0.24 & -0.28 \\
    \hline
    $\omega_b$ & $0.02235 \pm 0.00015$ & -0.64 & -0.24 & 1.00 & 0.45 \\
    \hline
    $n_s$ & $0.9644 \pm 0.0043$ & -0.73 & -0.28 & 0.45 & 1.00 \\
    \hline
    \end{tabular}

    \caption{\textit{Top.} Pantheon+ SNe data compressed in data-driven measurements of $1/E(z)$. \textit{Middle.} DES-Y5 SNe compressed in data-driven measurements of $1/E(z)$. \textit{Bottom.} CMB compressed variables.}
    \label{tab:SNe_DES_CMB_compressed}
\end{table}

\clearpage

\bibliographystyle{JHEP}
\bibliography{biblio.bib}
\end{document}